\documentclass[aps,prd,twocolumn,showpacs,floatfix,preprintnumbers,amsfont,amsmath,amssymb,nofootinbib, superscriptaddress,longbibliography]{revtex4-2}
\usepackage{graphicx} 
\usepackage{amsmath,amssymb}
\usepackage{gensymb}
\usepackage[T1]{fontenc} 
\usepackage{lmodern}

\usepackage{orcidlink}

\usepackage{aas_macros} 

\usepackage{lineno,hyperref}
\hypersetup{
 colorlinks=true,
 linktoc=all,
 linkcolor=blue,
 citecolor=blue,
 urlcolor = blue}

\def\te{t_{\rm E}}
\def\dls{d_{\rm LS}}
\def\mmbh{M_{\rm MBH}}
\def\mmbhz{M_{{\rm MBH},z}}

\begin{document}

\title{Self-lensing of moving gravitational-wave sources\\ can break the microlensing crossing timescale degeneracy}
\author{Helena Ubach\,\orcidlink{0000-0002-0679-9074}}
\email{helenaubach@icc.ub.edu}
\affiliation{Departament de F\'{i}sica Qu\`{a}ntica i Astrof\'{i}sica (FQA), Universitat de Barcelona (UB),  c. Mart\'{i} i Franqu\`{e}s, 1, 08028 Barcelona, Spain}
\affiliation{Institut de Ci\`{e}ncies del Cosmos (ICCUB), Universitat de Barcelona (UB), c. Mart\'{i} i Franqu\`{e}s, 1, 08028 Barcelona, Spain}

\begin{abstract}
When a moving gravitational-wave (GW) source 
travels behind a massive astrophysical object, its signal is gravitationally lensed, 
showing a waveform distortion similar to a Paczy\'{n}ski curve. 
We present a first 
study 
on the lensing signature of a massive black hole (MBH)
on a frequency-dependent GW signal from a moving compact binary merger (CBC) source, focused on ground-based GW detectors. 
For both light and GW sources in a 
Keplerian circular orbit around a MBH lens, 
the self-lensing geometry breaks the  
microlensing degeneracy in the Einstein radius crossing timescale $\te$. 
The duration of the curve ($2\te$) 
becomes independent on the MBH mass $\mmbh$, and 
provides 
a direct value of 
the orbital distance $\dls$ of the source around the MBH.  
However, $\mmbh$ remains unknown. 
In GW signals, 
the redshifted mass $\mmbhz$ can additionally
be analytically
inferred from the 
interference pattern, 
once the modulation period $T$, the GW frequency $f$, and $\te$ are known:
\mbox{$\mmbhz\simeq 2.5\times 10^6\,M_\odot\,(\te/[100\,{\rm s}])\,(f\,T)^{-1}$}. 
If this lensing signature is not considered,  
it may be confused with other waveform distortions, especially in the modeling of overlapping CBC signals in next generation ground-based GW detectors. 
The observation of one of these curves and its associated parameters may help 
(1) constrain  
the orbital distance $\dls$ 
of 
sources, 
especially around low-mass MBHs at the center of star clusters and galaxies, (2) additionally estimate the mass $\mmbhz$ of these MBHs, and (3) infer 
the  
orbital inclination 
of the binary. 
Simultaneously 
obtaining  
$\dls$ and $\mmbhz$ through self-lensing can help constrain the astrophysical environments where  
GW signals  
come from.  
\end{abstract}

\maketitle

\section{Introduction}
 
When light and gravitational waves (GWs) travel through 
the space-time curvature of 
massive astrophysical objects, 
they are deflected and distorted by gravitational lensing. 
While lensing effects have already been detected on light \cite{walsh-79,alcock-93,udalski-93,bond-01,tyson-90}, 
 they have not been confidently detected on GWs yet 
\cite{hannuksela-19,dai-20,LVK-O3a-lensing, LVK-O3ab-Lensing,janquart-23,LVK-O4a-Lensing} (though there is a recent, highly ranked candidate \cite{231123,goyal-25,chan-25-231123,hu-25,26-chakraborty,25-shan}). 
GWs from binary mergers, usually referred to as compact binary coalescences (CBCs), have been detected by ground-based detectors from the LIGO-Virgo-KAGRA (LVK) Collaboration. To date, more than $300$ GW event candidates have been detected \cite{LIGO18-GWTC1,LIGO20-GWTC2,LIGO21-GWTC3,venumadhav-20,zackay-21,wadekar-23,LIGO25-GWTC4,LIGO26-GWTC5}.
As the number of GW events from CBCs increases, the lensing effect is expected to be detected on GWs soon \cite{ng-18,li-18,oguri-18,xu-ezquiaga-holz-22,mukherjee-21,wierda-21}. 

The approaches to the observation of lensing of light and lensing of GWs are generally different. 
For light,
lensing effects appear as either 
(1) multiple images of the same source, resolved through imaging (strong lensing \cite{zwicky-37,walsh-79}); (2) 
transient increase 
of the source brightness following a Paczy\'{n}ski curve, when the source temporarily travels behind the lens \cite{chang-refsdal-79,paczynski-86,alcock-93,udalski-93,bond-01}, seen through photometry (microlensing); (3) astrometric microlensing \cite{paczynski-95,hog-95,miyamoto-yoshii-95,walker-95,dominik-sahu-00}, usually simultaneous to microlensing and seen as the displacement of the centroid of light in the sky; (4) weak lensing, seen as coherent distortions of the background galaxies behind a mass distribution \cite{tyson-90,kaiser-93}. For transient GWs detectable in ground-based detectors, 
the equivalent of strong lensing 
is the detection of 
multiple images of a static source in time domain \cite{wang-96}. Given the transient nature of the detected compact binary merger (CBC) sources, the 
equivalent to microlensing by low mass lenses
are wave optics distortions \cite{nakamura-98,takahashi-03}, mostly studied on static sources. 
The equivalent of 
the Paczy\'{n}ski curve is usually not considered for transient GWs from CBCs in ground-based detectors. 
Studies either consider a static source and varying frequency \cite[e.g.,][]{liao-19,sun-fan-19,hou-20,biesiada-harikumar-21} or a moving source and a constant frequency, 
the latter of which has been analyzed for continuous GW sources in ground-based \cite[e.g.,][]{depaolis-01,basak-23,guo-24,sudhagar-25,yang-25,li-25} and space-based \cite{pijnenburg-24,santos-25} GW detectors. 
Studies considering joint time and frequency evolutions have been done
for space-based detectors \cite{itoh-09,dorazio-loeb-20,yu-21,yang-25}. To the best of the author's knowledge, no previous studies have been done for a frequency-dependent moving source in ground-based detectors. 

The frequency-dependent moving GW source is a case that can already be potentially detectable in nominal LVK frequencies, $(10-10^4)\,{\rm Hz}$ \cite{16-martynov}. It may 
become especially relevant for future ground-based detectors such as Einstein Telescope (ET) and Cosmic Explorer (CE), when the signals are expected to be detected at lower frequencies, down to $\sim 3\,{\rm Hz}$ \cite[e.g.,][]{branchesi-23}, and therefore to last up to $\sim{\rm hours}$ in the initial inspiral phase of a CBC. The lensing signature of the moving source ---a Paczy\'{n}ski-like curve---  
on the CBC GW signal
may reach a  
comparable amplitude to 
the merger phase amplitude. 
The transient modification of the signal due to the lensing signature could lead to misinterpretations if this effect is not taken into account, especially when treating overlapping signals.

The observation of the lensing curve can give us information of the parameters of the lensing system. 
In microlensing of light, the duration of the curve (twice the Einstein radius crossing timescale $\te$) usually has a degeneracy between the lens mass, a combination of the distances, and the relative velocity of the lens and the source \cite[e.g.,][]{narayan-bartelmann-96,lee-17,gould-19}. The degeneracy may be broken by measuring the parallax \cite{refsdal-66,gould-94b,gould-00}, simultaneous parameters \cite[][for a review]{lee-17,gould-19}, finite-source effects \cite{gould-94a,nemiroff-94},
or relativistic effects \cite{rahvar-20}. 
By contrast, when the source is orbiting a massive black hole (MBH) that acts as the lens 
(\textit{self-lensing}), the Einstein radius crossing timescale is only dependent on the orbital distance $\dls$ from the source to the lens \cite{rahvar-11}.
For both light and GWs, the breaking of the Einstein crossing timescale degeneracy thus provides an independent way to measure the orbital distance $\dls$ of the source. 

Gravitational waves  
offer an additional opportunity 
to extract more parameters from the signal: their coherence and long wavelengths 
produce wave optics effects\footnote{There are a few special phenomena in light which may also exhibit coherent wave effects and the possibility to measure the lens mass from them \cite{katz-18,montero-camacho-19,jung-kim-20,sugiyama-20,jow-20,katz-20,paynter-21}.}, such as interference between the multiple images. 
The modulation of the signal by the interference pattern carries the information on the redshifted mass of the MBH $\mmbhz$. 
The modulations generally 
depend on the product between the lens mass $\mmbhz$ and the dimensionless source position $y$. 
In the static source case, $y$ should be measured independently to retrieve the mass $\mmbhz$. 
In the moving source case, as we will derive, 
$\mmbhz$ can be obtained 
through 
the crossing timescale $\te$ instead, which already implicitly involves the source position.

The combination of these capabilities for GW sources  
provides a way 
to potentially infer the orbital distance $\dls$
and the 
lens mass $\mmbhz$ 
simultaneously. 
The observation of  
these lensing curves and their associated orbital parameters  
may help constrain  
the distance distribution  
of CBCs, especially around low mass MBHs. It can also help disentangle scaling degeneracies between the orbital distance and the MBH mass which otherwise require the simultaneous measurement of the velocity and the acceleration of the source \cite{samsing-25-AGN}.

In this work, we obtain the lensing signature on a GW signal ``chirp'' from a quasi-circular CBC 
(Sec.~\ref{sec:signature}). We then analyze how to extract the orbital distance $\dls$ (Sec.~\ref{sec:distance}), the MBH mass $\mmbhz$ (Sec.~\ref{sec:mass}) and potentially the inclination of the binary (Sec.~\ref{sec:inclination}) from the signal features. We finally discuss the limitations of this study and future directions.

\section{Lensing curve on a chirping signal}
\label{sec:signature}

\begin{figure}[t]
\centering
\includegraphics[width=\linewidth]{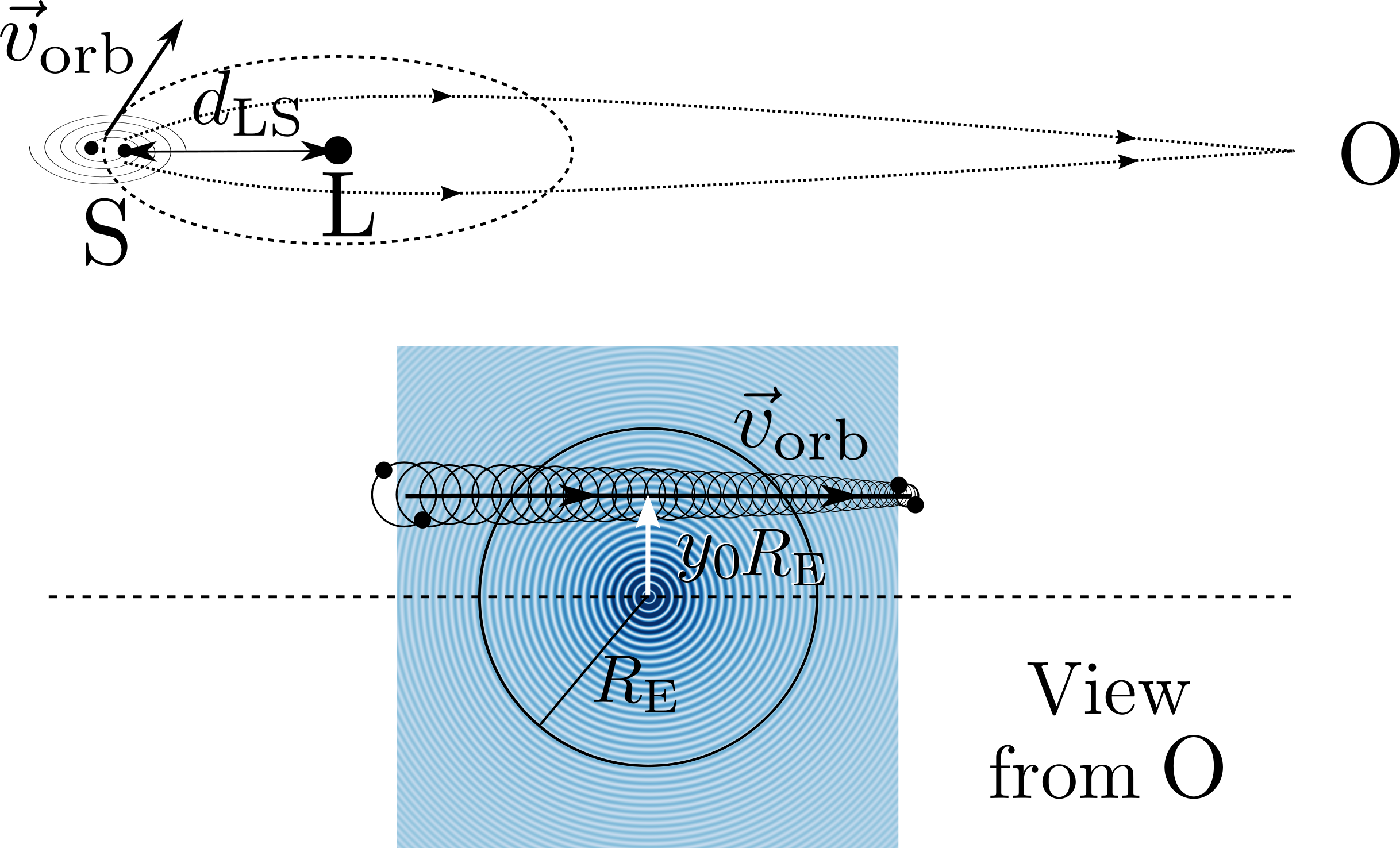}
\caption{
Illustration 
of the lensing configuration considered in this study. \textit{Top}: The GW source (S) is a compact binary orbiting a MBH at a distance $\dls$. The MBH acts as a gravitational lens (L), affecting the GW signal, with arrives distorted at the observer (O) when S, L and O are close to alignment. Not to scale. \textit{Bottom}: View from the observer, where the source travels behind the lens. The lensing effect is dominant behind the cross-section of Einstein radius $R_{\rm E}$ and, as we will see in Sec.~\ref{sec:mass}, the amplitude of the signal is modulated by the transmission factor ($F$ in Eq.~\eqref{eq:imprint-freq}) shown by the blue oscillating pattern here. The source can have an offset $y_0 R_{\rm E}$ with respect to the perfect alignment (dashed line going through the center of the lens), where $y_0$ is the dimensionless source position at closest approach.}
\label{fig:geometry}
\end{figure}

The astrophysical  
configuration 
considered in this work 
is shown in Fig.~\ref{fig:geometry}: a compact binary 
at S orbits a massive black hole (MBH) of mass $\mmbhz$ at L, at an orbital distance $\dls$. The compact binary is a source of GWs and results in a CBC. In this work, we focus on GW frequencies detectable by ground-based GW detectors: the inspiral is described in Appendix \ref{sec:appendix-chirp}, and is ruled by the ``chirp mass'' $M_{\rm chirp}$ of the source, a combination of the binary components masses defined in Eq.~\eqref{eq:chirp-mass}.  When the source travels behind the MBH, the emitted GWs undergo gravitational lensing by the MBH lens.  
When the binary is bound to the MBH's potential, we refer to the case as \textit{self-lensing} \cite{ubach-25}, and the velocity of the source corresponds to the orbital velocity $v_{\rm orb}$. 

The source travels behind the lensing cross-section, generally\footnote{Following classical microlensing conventions, we are taking a maximum source position of $y_{\rm max}=1$ as the cross-section threshold. 
For GWs, the detectability thresholds for $y$ can be different (usually $\gtrsim 1$) than for electromagnetic lensing, and in that case the cross-section is $\pi R_{\rm E}^2 y_{\rm max}^2$.}
given by $\pi R_{\rm E}^2$, where $R_{\rm E}=\sqrt{2 R_{\rm S} \dls d_{\rm L}/d_{\rm S}}$ is the Einstein radius of the lens, which for self-lensing is simplified to $R_{\rm E}\simeq\sqrt{2 R_{\rm S} \dls}$ \footnote{Throughout this work, due to self-lensing, the distance $d_{\rm L}$ from the observer to the lens can be taken as practically equal to the distance $d_{\rm S}$ from the observer to the source.},  
where $d_{\rm L}$ and $d_{\rm S}$ are the distance from the observer to the lens and to the source respectively, and $\dls$ the distance between the lens and source, which here coincides with the orbital distance. $R_{\rm S}=2 G \mmbh/c^2$ is the Schwarzschild radius of the MBH of mass $\mmbh$, where $G$ is the gravitational constant and $c$ is the speed of light in vacuum.
The lens mass $\mmbh$ used here is the intrinsic mass. Later in the manuscript we will use the redshifted mass \mbox{$\mmbhz\equiv M_{\rm MBH}(1+z)$} for a lens at a cosmological redshift $z$.

The source is generally not perfectly aligned (i.e., not crossing exactly behind the lens), it has a physical offset $\eta_0 \simeq y_0 R_{\rm E}$ 
where $y_0$ is the dimensionless source position (``impact parameter'', sometimes known as $u_0$)
of closest approach to the lens 
at time $t_0$. During the time of crossing, we consider a constant transverse velocity and the thin lens approximation. As the source moves, the source position $y$ with respect to the lens changes over time ($t$),
\begin{align}
y(t)&= \sqrt{y_0^2+(t-t_0)^2/\te^2}~, 
\label{eq:y-t}
\end{align}
where $\te$ is 
the Einstein radius crossing timescale \mbox{$\te=R_{\rm E}/v_{\rm orb}$}.

\subsection{The crossing timescale depends directly on the orbital distance}

In self-lensing \cite{beskin-tuntsov-02,rahvar-11,kasuya-11,han-16}, the orbital velocity of the source, if assumed Keplerian and circular\footnote{This assumption is taken because we focus on self-lensed GW mergers: 
most of them are expected in AGN disks \cite{gondan-kocsis-22,leong-25,ubach-25}, where the gas tends to practically circularize the orbits of embedded objects over time \cite[e.g.,][]{trani-di-cintio-25}. 
For the eccentric orbit case (for either light or GW sources), we refer the reader to Appendix \ref{sec:appendix-eccentric}. There, we 
also show that $\te$ remains in the same order of magnitude as the circular case for $e\lesssim 0.5$.
}, is $v_{\rm orb}\simeq \sqrt{G\mmbh/\dls}$. In the crossing timescale $\te$, the increase in orbital velocity for a larger $\mmbh$ is compensated by a larger 
cross-section of radius $R_{\rm E}\simeq\sqrt{2 R_{\rm S} \dls}$ to cross, 
which results on a crossing timescale only dependent on $\dls$ \cite{rahvar-11,dorazio-distefano-20}:
\begin{equation}
\te = \frac{2 \dls}{c}~.
\label{eq:tcross}
\end{equation}
Compared to a general source ---where $\te$ depends on the combination of the distances $\dls d_{\rm L}/d_{\rm S}$, on the relative velocity between lens and source, and on the mass of the lens---, this result 
is not degenerate and provides a direct value of the distance $\dls$, as we analyze in the following.

\subsection{Lensing signature on the GW waveform}

In frequency domain, the lensed GW strain $\tilde{h}(f)$ that describes the deformation of space-time by GWs can be obtained as 
\begin{equation}
\tilde{h}(f)= \tilde{h}_{\rm UL}(f) F(f,y)~,
\label{eq:imprint-freq}
\end{equation}
where $\tilde{h}_{\rm UL}(f)$ is the unlensed strain of a GW signal. The GW signals are modeled with waveforms, which are described through the strain.
In general, the transmission factor $F(f,y)$ that imprints the lensing effect depends both on the GW frequency $f$ and on the source position $y$ introduced in Eq.~\eqref{eq:y-t}. When the source is moving, both these variables change over time, $f(t), y(t)$, producing a complex situation. To find an analytical description, we use the point mass lens (PML) model, which is suitable for compact lenses such as our MBH case.

To obtain the lensed waveform in time domain, $h(t)$, we need to do the inverse Fourier transform of $\tilde{h}(f)$,
\begin{equation}
h(t) = \int df \, \tilde{h}(f) e^{i2\pi ft} = \int df \, \tilde{h}_{\rm UL}(f) F(f,y(t)) e^{i2\pi ft}.
\label{eq:fourier1}
\end{equation}

Our astrophysical situation allows us to use the Geometrical Optics (GO) approximation. 
For the approximation to be valid, the source position $y$ needs to fulfill the condition \cite{bulashenko-ubach-22} 
\begin{equation}
y\gtrsim 0.0125 \left(\frac{10\,{\rm Hz}}{f}\right) \left(\frac{10^5\,M_\odot}{\mmbhz}\right)~,
\label{eq:condition-GO}
\end{equation}
which will be met for the values considered in this study, usually a redshifted lens mass $\mmbhz\gtrsim10^5\,M_\odot$, and GW frequencies $f\gtrsim10\,{\rm Hz}$.

The GO approximation allows us to simplify the transmission factor $F$ in Eq.~\eqref{eq:imprint-freq}, and thus reduce the strain of the GW signal to a sum over different images of magnification $\sqrt{\mu_{\rm i}}$ each, 
\begin{align}
\tilde{h}(f) &= \tilde{h}_{\rm UL}(f) \sum_{\rm images (i)} \sqrt{\mu_{\rm i}} e^{i2\pi f t_{\rm i}-i\pi n_{\rm j}/2} \\
&= \tilde{h}_{\rm UL}(f) e^{i2\pi f t_1} \left[ \sqrt{\mu_1}\,\, +\sqrt{\mu_2} e^{i2\pi f \Delta t-i\pi /2} \right]~,
\label{eq:h-lensed-GO}
\end{align}
where in the last step we assume the PML model, which has two images. Here, $t_1$ is the arrival time of the first image, and $\Delta t$
is the time delay of the second image with respect to the first image. 
The time delay for the PML is well approximated by 
\begin{equation}
\Delta t\simeq4 y \frac{R_{\rm S}(1+z)}{c} 
\label{eq:tdelay}
\end{equation}
 for $y\lesssim 0.5$ \cite{bulashenko-ubach-22}. The method described in this work could be extended to other central lens models that converge to the same expression for $\Delta t$. 
$n_{\rm j}$ is the topological Morse phase of each image. The GO 
approximation allows us to split the integral. 
Furthermore, in GO, the amplitude $|F|=\sqrt{\mu_{1,2}}$ of each image is independent of $f$, it only depends on $y$: 
\begin{equation}
    \sqrt{\mu_{1,2}} = \frac{1}{2} \left(\frac{y}{\sqrt{y^2+4}} +\frac{\sqrt{y^2+4}}{y}\pm 2\right)^{1/2}~.
\label{eq:mu-GO}
\end{equation}
The treatment of the integral gets simpler, because a phase shift in the frequency domain translates to a time shift in time domain. 

The time domain signal is finally described by 
\begin{align}
h(t)=& |\sqrt{\mu_1}\, h_{\rm UL}|\,\cos \{\Phi[t_{\rm coal}-(t+t_1)]\}  
\\
+& |\sqrt{\mu_2}\, h_{\rm UL}|\,\cos\{\Phi[t_{\rm coal}-(t+t_1)+\Delta t]-\pi/2\}
~,
\label{eq:h-t-lensed}
\end{align}
where $\Phi$ is the phase evolution of an unlensed GW chirp given in Eq.~\eqref{eq:phase-chirp}, which is written with respect to the time to coalescence $t_{\rm coal}$. 
Therefore, the signal is a combination of two images of the original source, scaled by a magnification $\sqrt{\mu_{1,2}}$ each and 
separated 
by a time delay $\Delta t$. For the moving lens, $\sqrt{\mu_{1,2}}$, $\Delta t$ and $t_1$ are functions of $y(t)$ [Eq.~\eqref{eq:y-t}] and change over time. 

To assess the significance 
of the lensing curve on the signal, we work with the following premise. 
The  
strain gives a sense of the intensity of gravitational waves $\sim h^2$. 
If the peak of the Paczy\'{n}ski curve has a comparable value to the amplitude at merger ($h_{\rm curve}\sim h_{\rm merger}$) and both frequencies are in-band, the intensity (excess power) of the lensing curve could be measured by the detectors, e.g. seen in spectrograms. However, a measure of excess power alone may not be sufficient to identify the lensing feature, and a more detailed future study is required to determine the detectability of the signal. 

\begin{figure*}
\includegraphics[width=\linewidth]{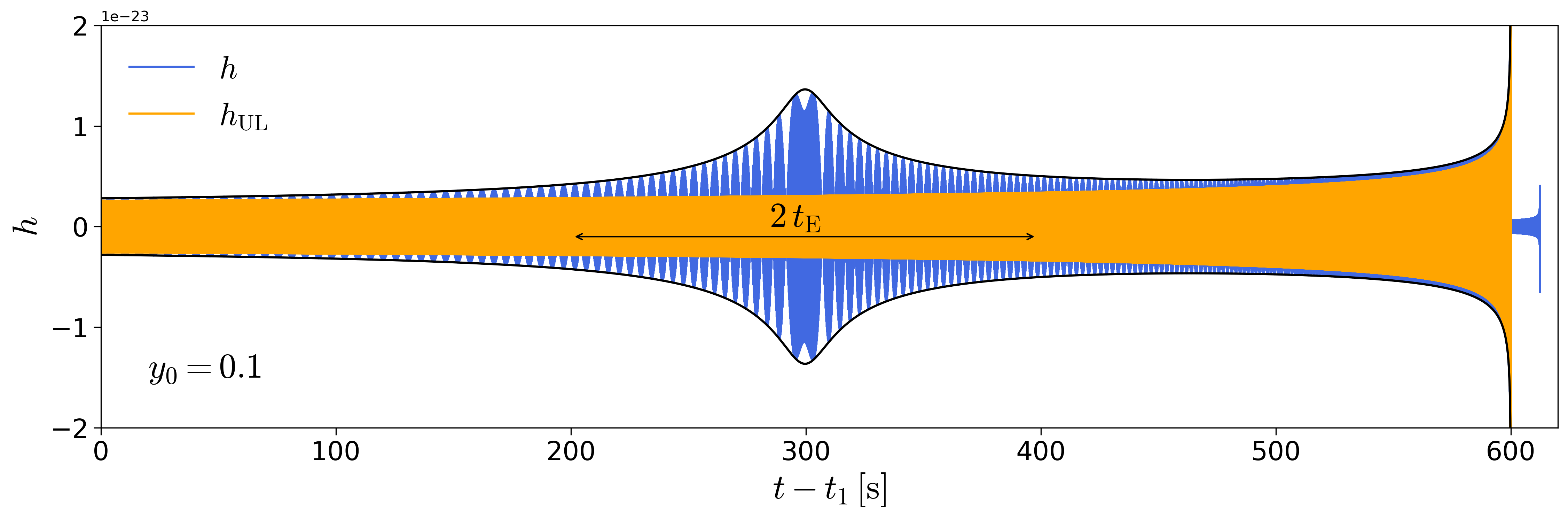}
\caption{Representation of a GW signal CBC ``chirp'', through the strain $h$ (relative deformation of space-time) as a function of time $t$. To compare it to the unlensed signal $h_{\rm UL}$, the lensed signal $h$ is represented in the time reference of the arrival time of the first image, $t_1$. The amplitude of the Paczy\'{n}ski-like curve on the inspiral phase of the signal can be comparable to (or even larger than) the final merger. The parameters used here are a source ``chirp mass'' $M_{\rm chirp}=5 \,{\rm M}_\odot$, a distance from the observer $d_{\rm L}=100\,{\rm Mpc}$ (just affecting the overall amplitude of the strain), $y_0=0.1$, $\mmbh=10^5\,{\rm M}_\odot$ and 
$\dls\simeq1.5\times10^{10}\,{\rm m}$ (0.1 $\rm AU$), corresponding to a distance of $50$ Schwarzschild radii from the MBH. In self-lensing, the 
width of the lensing curve ($2\te=4\dls/c$) is independent on $\mmbh$ [Eq.~\eqref{eq:tcross}]. Instead,  
as we will see in Sec.~\ref{sec:mass}, 
$\mmbh$ can be obtained from the interference pattern modulations on the signal 
(Fig.~\ref{fig:zoom-waveform}). 
We represent the inspiral part of the signal from $f\gtrsim 5\,{\rm Hz}$, with the future ET/CE in mind. For examples on more limited frequency ranges (for instance, for LVK), see Appendix \ref{sec:appendix-fmin}. 
}
\label{fig:lensed_inspiral_waveform}
\end{figure*}

A lensed CBC signal is shown in Fig.~\ref{fig:lensed_inspiral_waveform}. Its waveform is described in Appendix \ref{sec:appendix-chirp}. To interpret it, we start chronologically with the source far from alignment, when the first image looks unlensed ($\sqrt{\mu_1}\rightarrow1$), the second image is very faint ($\sqrt{\mu_2}\rightarrow 0$) 
and the time delay between them is large. As the source moves behind the lens, both 
images are magnified and the time delay shortens. At $t_0$, the images arrive simultaneously and at their maximum joint amplification, corresponding to the peak. As the source moves away from alignment, 
the 
initial situation is recovered, 
$\sqrt{\mu_1}\rightarrow 1$ and $\sqrt{\mu_2}\rightarrow 0$. The interference ceases when the first image of the merger is detected. The second image of the merger arrives alone after a time delay (for the source position at the time of merger).

\begin{figure}[h]
\begin{minipage}{\columnwidth}
\includegraphics[width=\columnwidth]{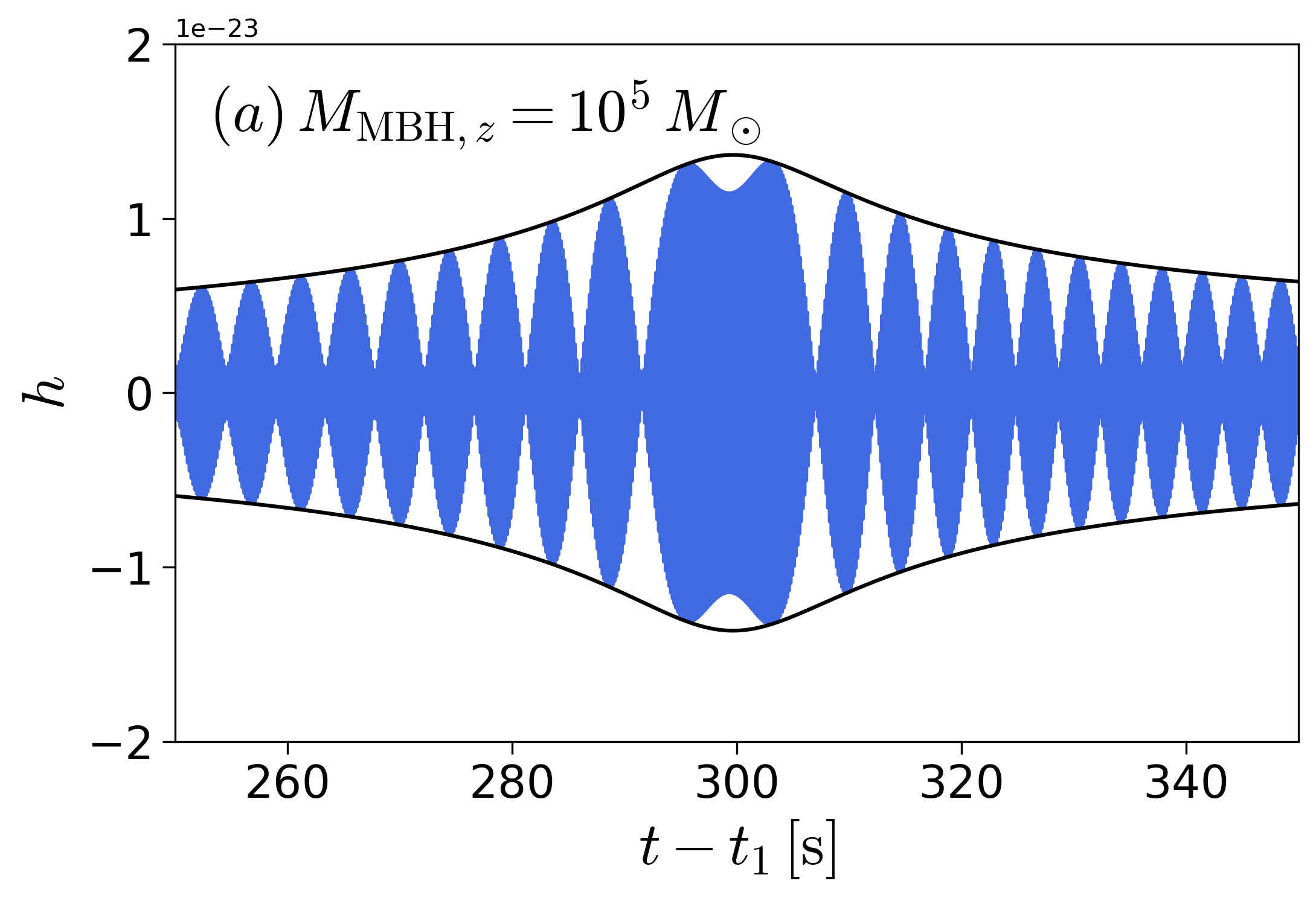}
\end{minipage}
\begin{minipage}{\columnwidth}
\includegraphics[width=\columnwidth]{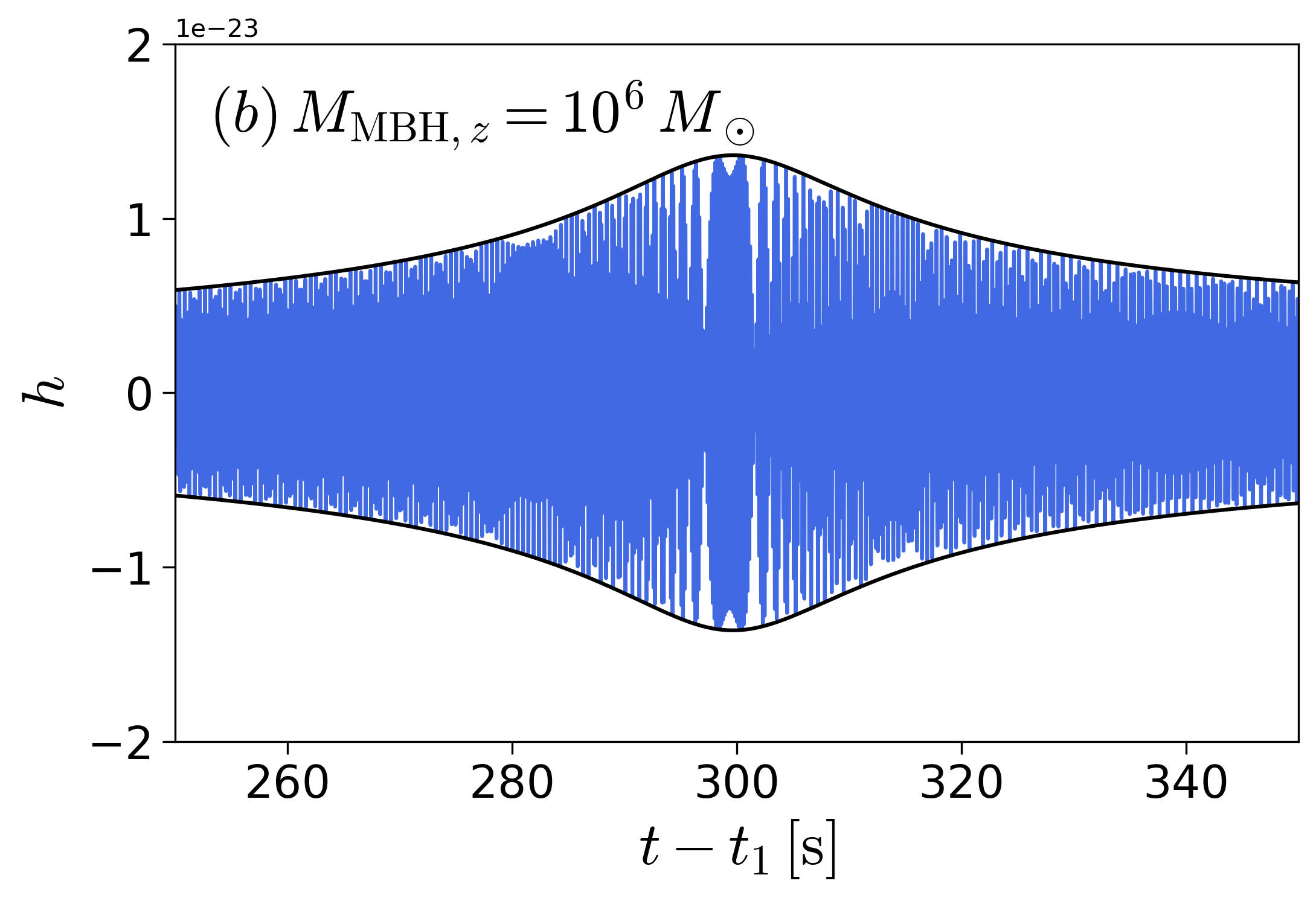}
\end{minipage}
\caption{Zoom-in view of Fig.~\ref{fig:lensed_inspiral_waveform}, at the peak of the lensing curve at GW frequencies 
$f\simeq 7\,{\rm Hz}$. 
The interference pattern modulating the GW signal
depends on the mass, (a) $\mmbhz=10^5\,M_\odot$, (b) $\mmbhz=10^6\,M_\odot$,  and the distance 
$\dls\simeq 1.5\times 10^{10}\,{\rm m}$ (or equivalently $\te\simeq 100\,{\rm s}$).
The modulation period far from the peak can be estimated with Eq.~\eqref{eq:modulation-period}.}
\label{fig:zoom-waveform}
\end{figure}

Although the source waveform $h_{\rm UL}(t)$ is chosen to be a CBC ``chirp'', with frequency dependence $f(t)\propto M_{\rm chirp}^{-5/8}$ \cite[e.g.,][]{maggiore-07}, the method can be applied to any smoothly varying source. 
A zoom-in view of Fig.~\ref{fig:lensed_inspiral_waveform} is shown in Fig.~\ref{fig:zoom-waveform}: while the interference pattern modulations are distinguishable, the GW oscillations have too many cycles to be resolved by eye. 
In this work, the amplitude already gives us enough information. However, the phase may give additional information about environmental phase shifts, which are not taken into account here and may be required in more detailed searches.   

In ground-based GW detectors, this lensing curve on the GW signal may appear as an excess power. If not taken into account, 
the full signal may be missed or misinterpreted. 
Although in most cases the lensing signature might not be confused with short bursts (glitches or other burst signals such as eccentric pulses/bursts \cite{samsing-25-GWpulsar}) because $\te$ is generally longer than their duration, the lensing curve should be taken into account when treating multiple overlapping signals in next-generation detectors. 

The analytical description of waveform relies on several assumptions, which we list here. For the geometry, we assume $\dls \ll d_{\rm S}, d_{\rm L}$, for a static lens and a static observer in a non-expanding Universe. The orbit is assumed to be circular, though wide enough to assume a transverse, non-accelerating, GW source motion (not accounting for the curvature of the orbit, because the Einstein radius crossing is in a small arc of the orbit). We also assume the thin lens approximation. The GW source is taken as the inspiral of a CBC (described in Appendix \ref{sec:appendix-chirp}) and point-like in nature. For the lensing effect, we consider the lens potential as a weak field, which can be described as a scalar field for propagation \cite[e.g.,][]{takahashi-03}. This scalar field is assumed (to the leading order) to have the same amplitude across the lens plane \cite[e.g.,][]{bulashenko-ubach-22}. The lens MBH in this work does not spin. Finally, the large lens mass allows us to use the GO approximation, while the distance to the caustic remains close enough to approximate the time delay as shown in Eq.~\eqref{eq:tdelay}. 
The simplification of the problem allows us to obtain analytical expressions for the system parameters, such as the lens mass from the interference pattern, later in Eq.~\eqref{eq:modulation-period-physical}.

\begin{figure}
\centering
\includegraphics[width=\linewidth]{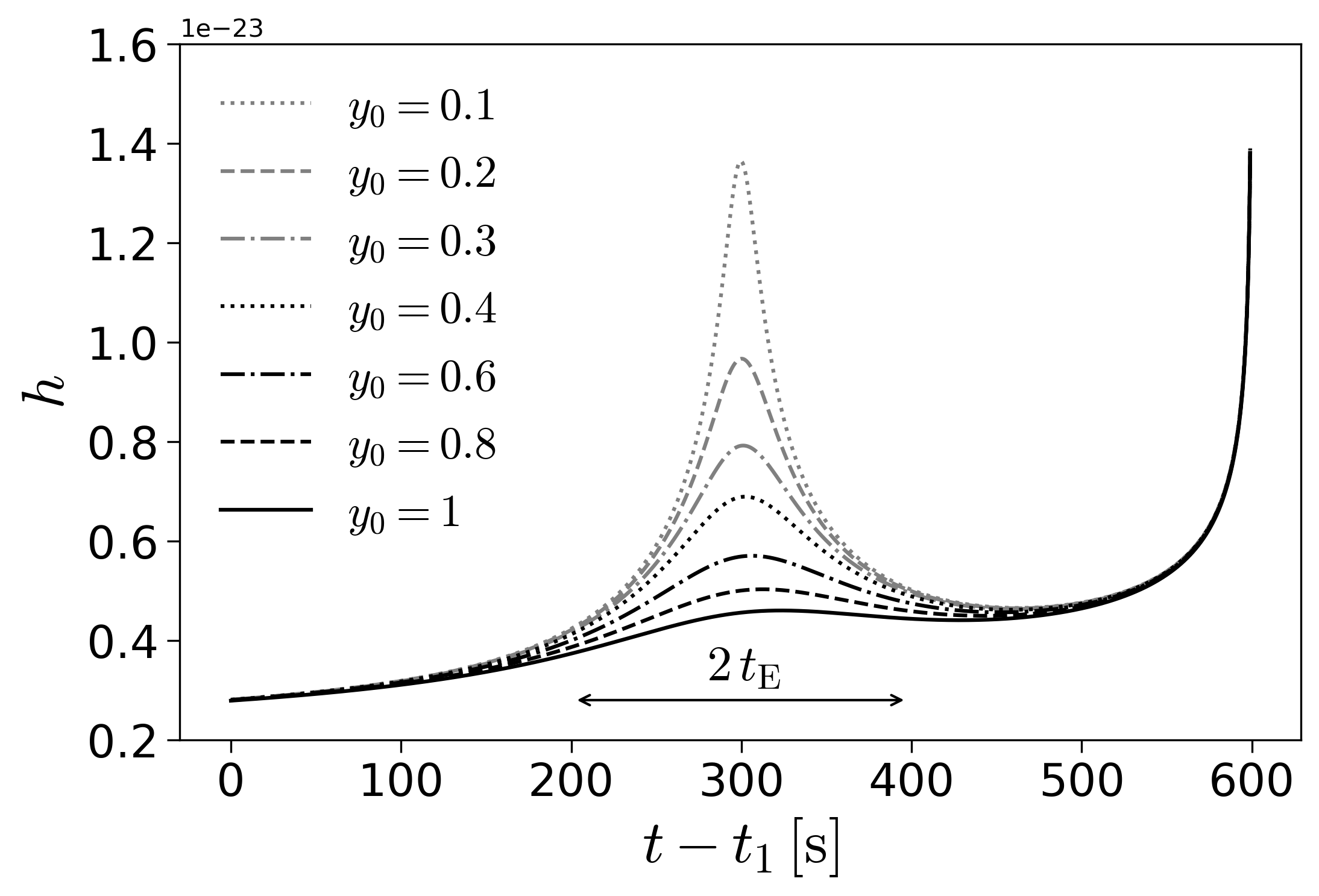}
\caption{Amplitude of the GW signal (envelope curve) as a function of the closest source position $y_0$. The curve follows a Paczy\'{n}ski-like curve. The width of the curve is $2\te$: it is practically independent on $y_0$, and for self-lensing it only depends on $\dls$ [Eq.~\eqref{eq:tcross}].} 
\label{fig:paczynski}
\end{figure}

\section{Obtaining the orbital distance}
\label{sec:distance}

The envelope of the lensing signature follows a Paczy\'{n}ski-like curve \cite{paczynski-86} in the part of the inspiral where the source passes behind the lens. 
By contrast to the optical microlensing curve, given by multiplying the signal by the amplification $\mu_1+\mu_2=(y^2+2)/(y\sqrt{y^2+4})$ \cite{paczynski-86}, the GW lensing curve depends on $\sqrt{\mu_1}$ and $\sqrt{\mu_2}$ multiplying each of the time-dependent images. 

The width of the curve is given by $2\te$. Figure \ref{fig:paczynski} shows that this width is practically independent on $y_0$. 
Therefore, 
the distance $\dls$ can then directly be obtained by measuring the crossing time $2\te$ [Eq.~\eqref{eq:tcross}]:
\begin{equation}
\dls\simeq 1.5\times 10^{10}\,{\rm m}\,\left(\frac{\te}{100\,{\rm s}}\right)~.
\label{eq:distance-physical}
\end{equation}

The prospects for observability depend on the astrophysical nature of the sources that can emit light or GWs:

\textbf{Observability for sources of light}. 
The light-curve may be seen for long timescales, provided that the source is at a distance $\dls$ larger than the tidal disruption radius of the source $R_{\rm tidal}$. 
For main sequence stars, the distance to the central massive BH to avoid tidal disruption should be $\dls \gtrsim 10^{13}\,{\rm m}\,[\mmbhz/(10^6\,M_\odot)]$, assuming the scaling relation $R_\star/R_\odot\simeq(M_\star/M_\odot)^{0.8}$ \cite[e.g.,][]{kippenhahn-weigert-90, chen-kipping-17}.
For typical  
$R_{\rm tidal}$ 
of stars around MBHs, we can expect $\te\gtrsim 3\,{\rm hours}$. 
Low-mass MBHs  
may be favored due to their potential to produce short timescales 
because \mbox{$\dls>R_{\rm tidal}\propto \mmbh^{1/3}$}, which can be interesting for the MBH at the center of our Galaxy (expanded in the Discussion, \ref{sec:discussion}) or intermediate-mass black holes (IMBHs) at the centers of star clusters. 

\textbf{Observability for GW signals}. 
The full lensing curve can be seen when $2\te<t_{\rm insp}$, where $t_{\rm insp}$ is the inspiral time given by \cite{maggiore-07}
\begin{align}
t_{\rm insp} &= \frac{5}{256}\, \left(\pi f_{\rm th}\right)^{-8/3}\left(\frac{G M_{\rm chirp}}{c^3}\right)^{-5/3}\\
&\simeq 6 \times 10^2 \,{\rm s}\,\left(\frac{5\,{\rm Hz}}{f_{\rm th}}\right)^{8/3}\left(\frac{5\,M_\odot}{M_{\rm chirp}}\right)^{5/3}~.
\end{align}
Here $M_{\rm chirp}$ is the ``chirp mass'' of the source defined in Eq.~\eqref{eq:chirp-mass}, and $f_{\rm th}$ is the threshold frequency 
from where the signal starts to be detectable. 
The condition $2\te<t_{\rm insp}$ leads to
\begin{equation}
\dls\lesssim 5\times10^{10}{\rm m}
\left(\frac{5\,{\rm Hz}}{f_{\rm th}}\right)^{8/3}
\left(\frac{5\,M_\odot}{M_{\rm chirp}}\right)^{5/3}~.
\label{eq:dls-condition}
\end{equation}
Otherwise, only part of the curve is observable. For ET, with a fiducial lower frequency of $f_{\rm th}\sim3\,{\rm Hz}$, one may take 
$f_{\rm th}=5\,{\rm Hz}$. For LVK, one should consider larger threshold frequencies than the nominal lower frequency $f_{\rm th}\sim10\,{\rm Hz}$.

Such close distances may be reached in active galactic nuclei (AGN) disks, if the sources are in migration traps \cite{mckernan-12,bellovary-16,secunda-19}. 
Migration traps are specific orbits where stellar-mass BHs and other objects may get trapped due to an equilibrium of gaseous torques. Therefore, they are a promising site for these objects to accumulate and form binaries. The precise values of their radii are an active area of research, though they are expected to scale with the Schwarzschild radius of the central MBH and be in the range $\sim(10-1000)\,R_{\rm S}$.
In order to see the full curve, as derived in Eq.~\eqref{eq:dls-condition}, the MBH mass has to be lower than a threshold, because of the scaling $\te \propto \dls\propto R_{\rm S}\propto \mmbh$. For $M_{\rm chirp}=5\,M_\odot$, $f_{\rm th}=5\,{\rm Hz}$, if we fix $\dls\sim 50\,R_{\rm S}$, a complete curve appears when $\mmbh\lesssim 3\cdot 10^5\,M_\odot$, while for the close $\dls\sim 10\,R_{\rm S}$  \cite{peng-21,leong-25}, it appears for $\mmbh\lesssim 2\cdot 10^6\,M_\odot$. Consequently, the detectability of the whole lensing curve in AGN disks may be biased towards low-mass MBH lenses.

\section{Obtaining the MBH mass}
\label{sec:mass}

The lensed GW signal presents a modulation of the amplitude 
which can be seen in Figs.~\ref{fig:lensed_inspiral_waveform},~\ref{fig:zoom-waveform}. 
The modulation is caused by
the phase difference between the lensed images, which creates an interference pattern that can also be found in continuous GW signals \cite{basak-23,guo-24,sudhagar-25,yang-25,li-25,sudhagar-25}. In CBC signals, the period of the modulations decreases as frequency increases.
The period of this modulation can be estimated in the GO approximation from the relative phase between the lensed images in Eq.~\eqref{eq:h-lensed-GO} 
 \cite{savastano-24}, from equating $e^{i2\pi f \Delta t}\equiv e^{i 2 \pi (t/T)}$.
The period is then $T=t/(f\Delta t)$.
$T$ can be written in terms of the Einstein cross-section $\te$, instead of $y$ (in the static lensing case), by
using 
$y\propto t/\te$. 
This relation is only formally valid for the radial spacing (interference pattern in Fig.~\ref{fig:geometry}).
Since in our case the source has an offset $y_0$, 
the period will asymptotically match the modulations 
far from the $t_0$ peak.
Nevertheless, the point of this estimation is to show that the period is inversely proportional to the mass of the MBH, 
\begin{align}
T&\simeq\frac{\dls}{2 R_{\rm S}}\, \frac{1}{1+z} \left(\frac{1}{f}\right) \\ 
&\simeq 2.5\,{\rm s}\left(\frac{\te}{100\,{\rm s}}\right)\left(\frac{10^5\,M_\odot}{\mmbhz}\right)\left(\frac{10\,{\rm Hz}}{f}\right) ~. 
\label{eq:modulation-period}
\end{align} 
By observing the modulation period $T$, the width of the curve $2\te$, and the GW frequency $f$ \footnote{The frequency is assumed to be directly measurable by the detector  from the incoming GWs.}, we can obtain the mass of the MBH:
\begin{equation}
\mmbhz\simeq 2.5\times 10^6\,M_\odot \left(\frac{\te}{100\,{\rm s}}\right)\,\frac{1}{f\,T}~.
\label{eq:modulation-period-physical}
\end{equation}

Although the mass $\mmbhz$ can also be extracted from the interference pattern in static lensing, 
it requires an independent measurement of the source position $y$.
For the moving source, instead, $\te$ becomes 
a 
much practical 
observable quantity in time domain: the width of the 
Paczy\'{n}ski-like curve.

For the astrophysical parameters we consider (typical MBH masses and LVK/ET/CE frequencies), most times 
the modulations in the interference pattern of the moving source
are expected to 
last $T\gtrsim {\rm ms}$. The 
interference pattern is
thus potentially resolvable in time domain in current ground-based GW detectors. 

The 
condition to observe at least a few of these modulations is that the 
period has to be shorter than the duration of the signal, $T<t_{\rm insp}$, leading to
\begin{equation}
\dls\lesssim10^4 \,R_{\rm S}(1+z)\left(\frac{f}{10\,{\rm Hz}}\right) \left(\frac{5\,M_\odot}{M_{\rm chirp}}\right)^{5/3}~,
\label{eq:m-condition}
\end{equation}
for $f_{\rm th}=5\,{\rm Hz}$.
These close orbital distances might be reached in AGN disks, where compact binaries may orbit in migration traps at \mbox{$\dls\sim(10-10^3)\,R_{\rm S}$} \cite{bellovary-16,secunda-19,peng-21,gangardt-24,grishin-24}.

\subsection{Joint observation of $\dls$ and $\mmbhz$}

\begin{figure}
\centering
\includegraphics[width=\columnwidth]{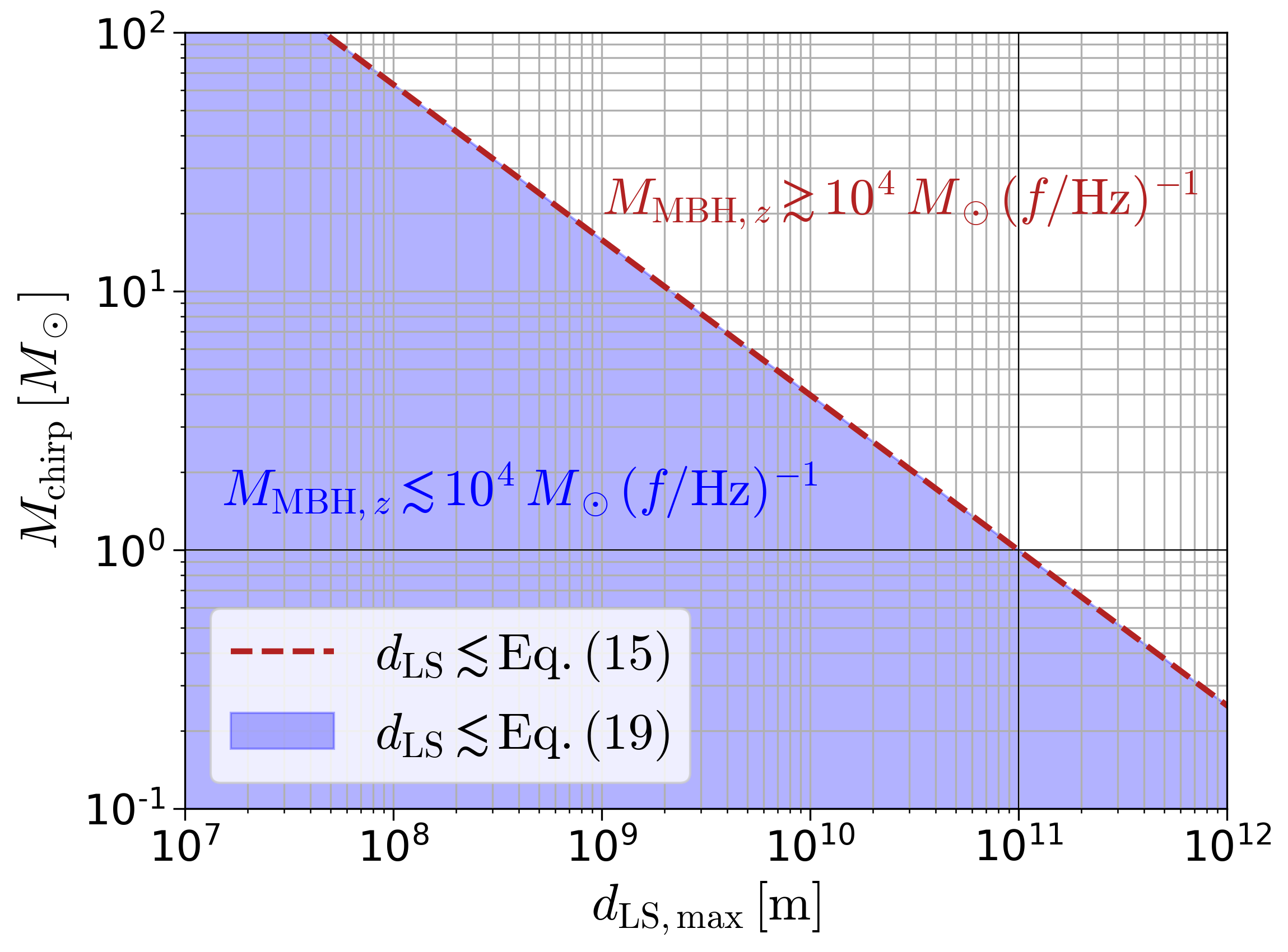}
\caption{Parameter thresholds to detect $\dls$ and $\mmbhz$ jointly. If $\mmbhz \gtrsim 10^4\,M_\odot\,({\rm Hz}/f)$, the maximum distance $\dls$ given a value of $M_{\rm chirp}$ is limited by observing $\te$, shown by the red dashed line (Eq.~\eqref{eq:dls-condition}). If $\mmbhz \lesssim 10^4\,M_\odot\,({\rm Hz}/f)$, the threshold lies in the blue region, where the exact curve depends on the different parameters in Eq.~\eqref{eq:m-condition}.}
\label{fig:constraints}
\end{figure}

To obtain $\mmbhz$ we require knowing $\te$ (from the width of the curve) and observing the modulations. 
For that, 
$\dls$ has to fulfill 
the two conditions in Eqs.~\eqref{eq:dls-condition} and \eqref{eq:m-condition}. For a mass above 
\begin{equation}
\mmbhz \sim 10^4\,M_\odot\,\left(\frac{\rm Hz}{f}\right)~,
\label{eq:limiting-m-joint}
\end{equation}
the limitation comes from being able to obtain $\dls$ from the crossing timescale, Eq.~\eqref{eq:dls-condition}, shown in a red dashed line in Fig.~\ref{fig:constraints}. For a lower $\mmbhz$ than Eq.~\eqref{eq:limiting-m-joint}, 
the limitation comes from being able to see at least a few modulations, Eq.~\eqref{eq:m-condition}, in the blue region in Fig.~\ref{fig:constraints}. Wave effects, in particular diffraction, are significant in the blue region: in fact, the value of the lens mass in Eq.~\eqref{eq:limiting-m-joint} corresponds to the threshold value of GO validity for $y\sim 1$ \cite[e.g.,][]{bulashenko-ubach-22}. In this region, the exact limiting curve depends on the various parameters in Eq.~\eqref{eq:m-condition}, and in general it is more restrictive on the maximum value of $\dls$ than Eq.~\eqref{eq:dls-condition}. 

From those joint constraints, 
we conclude that the threshold distances $d_{\rm LS, max}$ lie within observational ranges, comparing them to Schwarzschild radii values of $R_{\rm S}\simeq3\times 10^9\,{\rm m}\,(\mmbh/[10^6\,M_\odot])$. Therefore, it is possible to measure both $\dls$ and $\mmbhz$ in realistic astrophysical scenarios, especially in AGN disks. 

\begin{figure}
\centering
\includegraphics[width=\columnwidth]{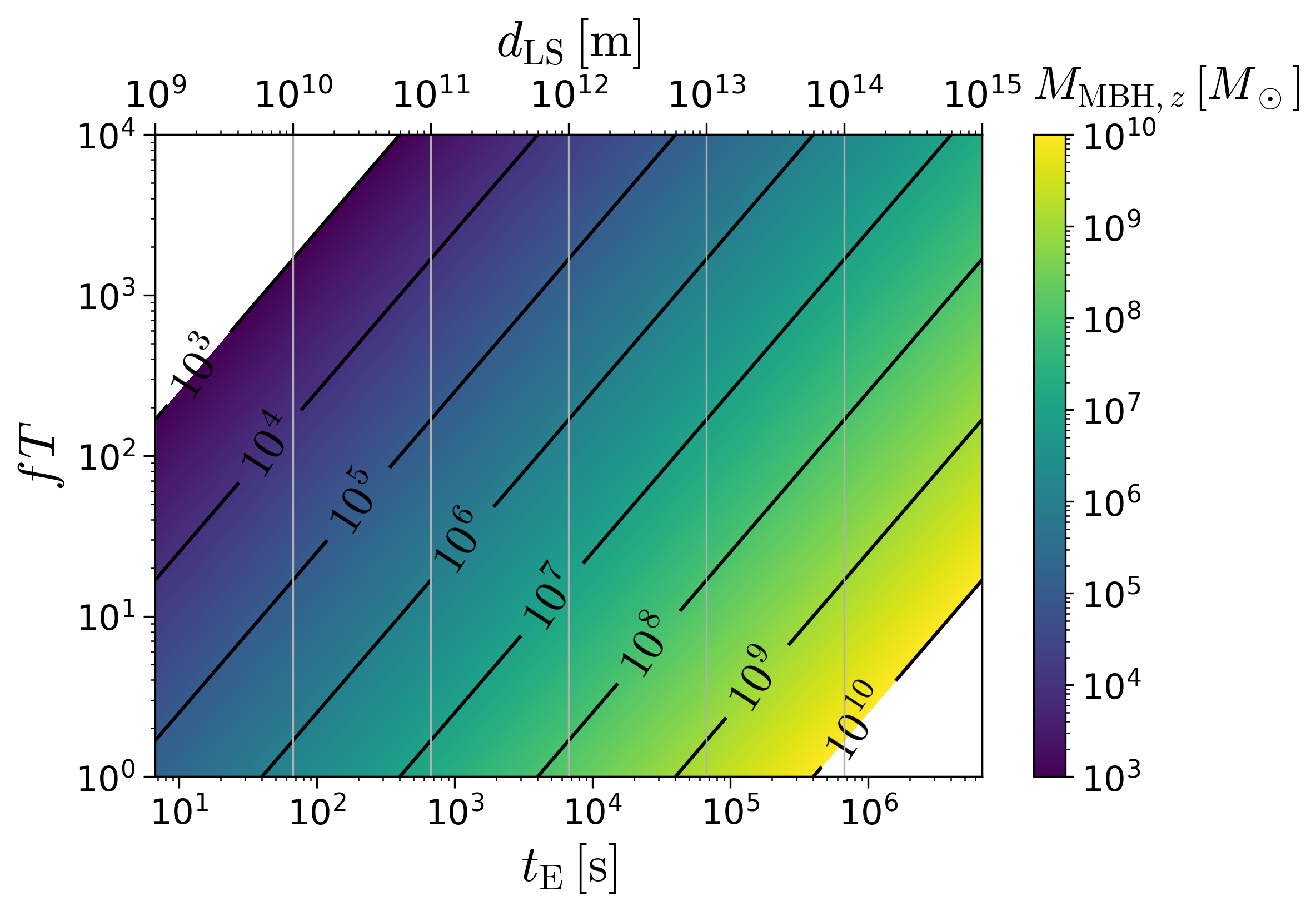}
\caption{Summary of the relations between parameters and how to obtain $\dls$ and $\mmbhz$ from measuring $\te$, $f$ and $T$. By measuring $\te$ from the width of the curve, we obtain $\dls$ through Eq.~\eqref{eq:tcross}. In GWs, the interference pattern can provide information about the modulation period $T$. By observing $T$, $f$ and $\te$, we can estimate $\mmbhz$ with Eq.~\eqref{eq:modulation-period-physical}.}
\label{fig:summary}
\end{figure}

Figure \ref{fig:summary} summarizes how to obtain simultaneously $\dls$ and $\mmbhz$ once $f$, $T$ and $\te$ can be measured from the GW signal.
The expected MBH masses and distance distributions depend on the specific astrophysical environment (star cluster, galactic nucleus, AGN disk) and its formation channels for GW sources. 
Measuring $\dls$ and $\mmbhz$ from the same signal may help distinguish the environment where the GW signal comes from, by comparing them to the expected values of $\dls$ and $\mmbhz$ for each environment \cite{samsing-25-AGN}.

\section{Inferring the orbital inclination of the binary}
\label{sec:inclination}

It may be possible to infer the inclination of the orbit $\theta$ with respect to the line of sight 
if the distance $\dls$, the MBH mass $\mmbh$, and $y_0$\footnote{The value of $y_0$ can be estimated from the relative amplitude of the curve, see Fig.~\ref{fig:paczynski}.} can be obtained from the signal. Given that $y_0\simeq\eta_0/R_{\rm E}$ is the normalized physical offset in self-lensing, and the inclination angle of the orbit with respect to the line of sight is given by $\theta\approx\eta_0/\dls$ for small angles, the inclination can be estimated as
\begin{align}
\theta &\approx y_0 R_{\rm E}/\dls \\
& \simeq 14\degree \, y_0 \left(\frac{\mmbh}{10^6\,M_\odot}\right)^{1/2} \left(\frac{10^{11}\,{\rm m}}{\dls}\right)^{1/2}~.
\label{eq:inclination}
\end{align}
Knowing the inclination of orbiting compact binaries can be interesting especially in AGN disk environments, 
which has recently been studied in \cite{dorazio-distefano-20,yu-21,postiglione-25,li-25} in the context of the space-based GW interferometer LISA, and is an area of active study in astrophysical dynamics \cite{rowan-25a}.

\section{Discussion}
\label{sec:discussion}

The prospects of observing the lensing Paczy\'{n}ski-like curve depend on the probability that the source is emitting the GW signal when it moves behind the lens. The probability of self-lensing of a static source can reach $\sim 2\%$ in AGN disks \cite{gondan-kocsis-22,leong-25,ubach-25}. In star clusters, self-lensing probabilities range between $\sim 10^{-8}-10^{-3}$ depending on the assumed formation mechanism of the BH binary. For details, we refer the reader to \cite{ubach-25}. The 
fraction 
of mergers from each formation channel and astrophysical environment is currently uncertain because, in many cases, the observed merger rates can be explained by each of these formation channels separately. The self-lensing probability quantifies the fraction of each channel's rate that would be affected by lensing. For the most promising and probable case in the AGN disk, expected BBH rates are \cite[e.g., most recent study,][]{rowan-25b} $0.73-7.1\,{\rm Gpc^{-3}\,{\rm yr^{-1}}}$ compared to the detected rate $14-26\,{\rm Gpc^{-3}\,{\rm yr^{-1}}}$ \cite{gwtc4-pop}, i.e. accounting from zero to half of all BBH mergers. 
If BBH mergers happened indeed in the tightest migration traps in AGN disks, self-lensing may affect up to $\sim 1\%$ of signals. Naturally, these estimates are dependent on the existence of BBH mergers in AGN disks, and the correct prediction of the radii of migration traps. Given that a moving source covers a larger region than a static source during the emission of the signal, it is expected that the probability is equal or higher than the static self-lensing case. However, it requires considering the distribution of signal durations as well. More work is needed to quantify it.

Wave effects appear here as the interference of the multiple images. The GO approximation holds for large masses, although the complete wave optics formalism may need to be considered for low masses (for example, IMBHs in star clusters) or very closely aligned orbits $y_0\rightarrow0$ \cite[e.g.,][]{sudhagar-25} when condition \eqref{eq:condition-GO} is not met. In this case, the integral in Eq.~\eqref{eq:fourier1} has a transmission factor that depends on both $f(t)$ and $y(t)$ (for example, in the PML, the product $f y$ appears as a variable), 
potentially preventing an analytical treatment, and requiring numerical computation of its waveform.

We have focused on the amplitude of the signal for this study, without entering into details of the phase. 
There is a global shift in Eq.~\eqref{eq:h-lensed-GO} given by the time dependence of the absolute time of arrival $t_1$ of the first image. The phase would need to be carefully modeled, especially if environmental effects that cause phase shifts \cite{takatsy-25} are taken into account. Precisely, Doppler shifts, gravitational redshift, aberration, beaming, absorption of radiation by the MBH and strong gravity effects in general \cite{santos-25} may become important at very short $\dls$ ($\sim R_{\rm S}$). 
We do not take these possible effects into account in this work. Nevertheless, a relation with the Doppler shift arises naturally from the time dependence of the time delay \cite{samsing-24b} \footnote{I thank and acknowledge  Johan Samsing and Miko\l aj Korzy\'{n}ski for insightful discussions on this point during the revision of the manuscript.}. We neglect the effect of the Earth's rotation on the amplitude.

Additionally, environmental effects may distort the signal or modify its inspiral time. The eccentricity of the binary can modulate the GW signal in a similar way to self-lensing: the modulations are nearly equally spaced in time domain, and their period decreases as frequency increases. The lens in this work is non-spinning, and the source is assumed to not be affected by precession. 
Further work is required to evaluate the potential degeneracies of these effects and disentangle them.

The excess power from the lensing curve may look like a (long) burst. Especially focusing in glitches, one might distinguish most of them from the Paczy\'{n}ski-like curve by their duration: most glitches are short, up to seconds, while the lensing curve is expected to last hundreds of seconds. Glitches can also be distinguished when they appear in a single detector and the GW signal appears simultaneously in more than one detector. Finally, the phase evolution can also be distinct in either case: the phase evolution in the lensing curve is smooth and predictable, while glitches do not generally have a smooth nor frequency-dependent phase evolution. Precisely in LVK observations, the phase evolution is analyzed with signal-consistency tests (e.g. $\chi^2$ test) to discard glitches whose phase differ from predictions for astrophysical GW signals.

Although we quantified the maximum $\dls$, some sources are limited by a minimum $\dls$ as well. For example, stars below $\dls<R_{\rm tidal}\simeq 10^{13}\,{\rm m}\,[\mmbhz/(10^6\,M_\odot)]$ are tidally disrupted, which reduces the probability of observing the optical Paczy\'{n}ski light-curve for systems with large $\mmbh$. 
For low MBH masses, the full lensing curve may still be detectable in observable timescales: for our Galaxy's central MBH, 
light-curves from orbiting stars may last for 
$2\te\gtrsim 8\,{\rm hours}$. 
However, currently known stars from the S-star cluster 
have a much larger orbital radius than the minimum $\dls$, 
and are not 
very likely to be microlensed by the MBH 
\cite{michalowski-21}, requiring very powerful telescopes \cite{bozza-mancini-12}. 

Even if the early part of the inspiral is in detectable frequencies, but below the threshold in strain amplitude (subthreshold signal), the enhancement of the GW amplitude by the lensing effect may bring part of the inspiral above the threshold, or accelerate the obtention of cumulative SNR. Both in this case and when the signal can be fully seen, the lensing Paczy\'{n}ski-like curve can act as an early warning of the subsequent merger.

This work can be extended to the inspiral part that is detectable by the future space-based GW detectors such as LISA, where it would be seen as a continuous source. Repeated lensing (around the MBH, as in \cite{dorazio-loeb-20,yu-21}) may be observed as repeated Paczy\'{n}ski curves, which may improve the determination of $\dls$ from their width. It may even be possible to track, through consecutive $\dls$ measurements, the dynamical evolution of the BBH over time.

\section{Conclusions}

The geometrical configuration in self-lensing 
breaks the usually found microlensing degeneracy for the Einstein crossing timescale $\te$.
It is therefore possible to extract the properties of the system from the lensing signature: 
\begin{itemize}
\item For both light and GW sources, the orbital distance $\dls$ of the source can be obtained analytically from the crossing timescale $\te$ (half width of the curve), \mbox{$\dls\simeq 1.5\times 10^{10}\,{\rm m}\,(\te/[100\,{\rm s}])$} [Eq.~\eqref{eq:distance-physical}]. For GW sources, the full curve may be observed for $\dls\lesssim 5\times 10^{10}{\rm m}\,(5\,M_\odot/M_{\rm chirp})^{5/3}$ when detectable frequencies are $\gtrsim 5\,{\rm Hz}$ (e.g., in ET). Short timescales are favored in systems with low $\mmbh$, which allow smaller values of $\dls$. For light, the light-curve can be observed for as long as the observation time.

\item The mass $\mmbhz$ of the MBH lens can be obtained from the interference pattern in GW signals. Differently from the static source case, $y$ is not directly required to infer the mass from the modulations: the mass can be obtained by knowing $\te$. Given the modulation period $T$, the instantaneous GW frequency $f$, and $\te$, \mbox{$\mmbhz\simeq 2.5\times 10^6\,M_\odot\,(\te/[100\,{\rm s}])\,(f\,T)^{-1}$} [Eq.~\eqref{eq:modulation-period-physical}]. These modulations are expected to be visible typically for \mbox{$\dls\lesssim 10^4\,R_{\rm S}\, (5\,M_\odot/M_{\rm chirp})^{5/3}$}. However, to measure $\mmbhz$, both conditions in Eq.~\eqref{eq:dls-condition} and \eqref{eq:m-condition} should be met. 

Figure \ref{fig:summary} summarizes how to obtain simultaneously $\dls$ and $\mmbhz$ by observing $f$, $T$ and $\te$. We conclude that 
$\mmbhz$ may be 
obtained  
at realistic distances in 
astrophysical environments through self-lensing. The mass measurement can be extended to other GW signals such as continuous GWs. 
    
Self-lensing therefore offers a complementary way to obtain $\dls$ and $\mmbhz$, 
compared to other ways such as
the simultaneous measure of the velocity and acceleration presented in \cite{samsing-25-AGN}. 

\item If the distance $\dls$, the mass $\mmbh$ and $y_0$ could be simultaneously obtained in GWs, it would be possible to infer the inclination of the orbit with respect to the line of sight, $\theta \propto  y_0 \mmbh^{1/2} \dls^{-1/2}$ [Eq.~\eqref{eq:inclination}].
\end{itemize}

Obtaining $\dls$ and $\mmbhz$ simultaneously would provide valuable information about the environment around the binary merger \cite[e.g.,][]{samsing-25-AGN}. 
Consequently, it would
help constrain the astrophysical environments where GW signals come from.

\section*{Acknowledgements}

This work has been inspired by 
conversations about moving sources with Sudhagar Suyamprakasam, Miko\l aj Korzy\'{n}ski and Sreekanth Harikumar at the workshop ``Lensing and Wave Optics in Strong Gravity'' at the Erwin Schr\"{o}dinger International Institute for Mathematics and Physics (ESI). I would like to thank the organizers of the workshop and its participants for such an inspiring atmosphere. I am grateful to Johan Samsing for engaging discussions later at the Niels Bohr Institute, and I would like to thank him and Lorenz Zwick for useful comments on early results of this work. I am deeply grateful for the hospitality of the Strong group and all colleagues at the Niels Bohr Institute, where part of this work was carried out. 
I would like to thank Mark Gieles and Jordi Miralda Escud\'{e} for providing useful feedback on the draft and for very insightful comments. I am further thankful to the anonymous referees, whose comments helped improve the clarity of this work, as well as to all colleagues that provided comments on the work.

Software: All the presented results are analytical, and the corresponding figures have been produced with Python (Numpy, Matplotlib). Fig.~\ref{fig:geometry} has been created with Inkscape. I declare that this work has been developed without use of artificial intelligence.

Financial support: FI-SDUR 2023 predoctoral grant (AGAUR, Generalitat de Catalunya), which also covered the visit to the Niels Bohr Institute. PID2024-159689NB-C22, CEX2024-001451-M funded by MCIN/AEI/10.13039/501100011033, and SGR-2021-01069 (AGAUR, Generalitat de Catalunya). 

\appendix

\section{Eccentric orbit}
\label{sec:appendix-eccentric}

In this work we have assumed a circular orbit of the BBH around the MBH. The motivation behind the circular orbit assumption is that the most promising environments (where we may find a GW source at a short orbital distance $\dls$, therefore a short $\te$) are AGN disks, where gas efficiently circularizes the orbit. However, for the general case, we need to consider the effect of orbital eccentricity, which can reintroduce a degeneracy.

\begin{figure}[h]
\centering
\includegraphics[width=0.5\columnwidth]{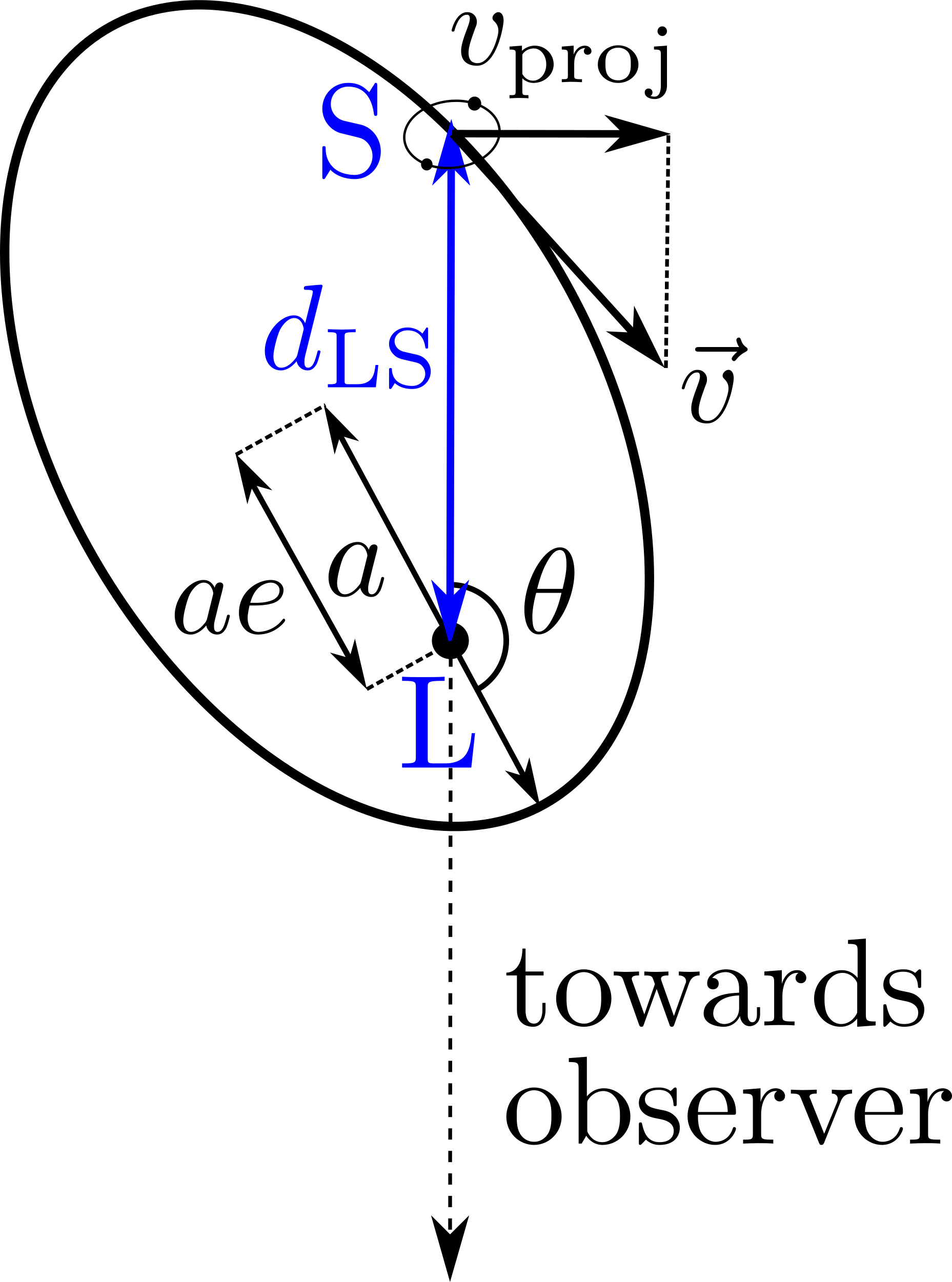}
\caption{Geometrical setup for an eccentric orbit of the BBH around the MBH. The inclination orbital plane is assumed to be practically edge-on for lensing to happen, reducing the dimensionality of the geometry to a plane. The orbital distance $\dls$ is dependent on the semi-major axis $a$, the eccentricity $e$, and the true anomaly $\theta$, as described in Eq.~\eqref{eq:dls-eccentric}.}
\label{fig:eccentric-orbit}
\end{figure}

Figure \ref{fig:eccentric-orbit} shows the geometrical configuration of the elliptical orbit \cite{kepler-1609-first-second-laws} in the presence of eccentricity. Because the orbital plane needs to be practically edge-on to observe self-lensing (the orbit needs to pass behind the lens as viewed from the observer), we can reduce the geometry to two dimensions. The orbital distance $d_{\rm LS}$ is then written as a function of the semi-major axis $a$, the orbital eccentricity $e$ and the orbital phase $\theta$ corresponding to the \textit{true anomaly} in orbital mechanics,
\begin{equation}
\dls(a,e,\theta) = \frac{a(1-e^2)}{1+e\cos\theta}~,
\label{eq:dls-eccentric}
\end{equation}
as the polar representation of the distance to the focus of the ellipse (position of the MBH).

The transverse velocity, 
shown as $v_{\rm proj}$ in Fig.~\ref{fig:eccentric-orbit}, 
can be obtained from conservation of angular momentum \mbox{$L=M\sqrt{GMa(1-e^2)}$} combined with Eq.~\eqref{eq:dls-eccentric}, yielding
\begin{equation}
v_{\rm proj} = \frac{L}{M d_{\rm LS}}=\frac{\sqrt{GM}(1+e\cos\theta)}{\sqrt{a(1-e^2)}}
\end{equation}

\begin{figure}
\centering
\includegraphics[width=\columnwidth]{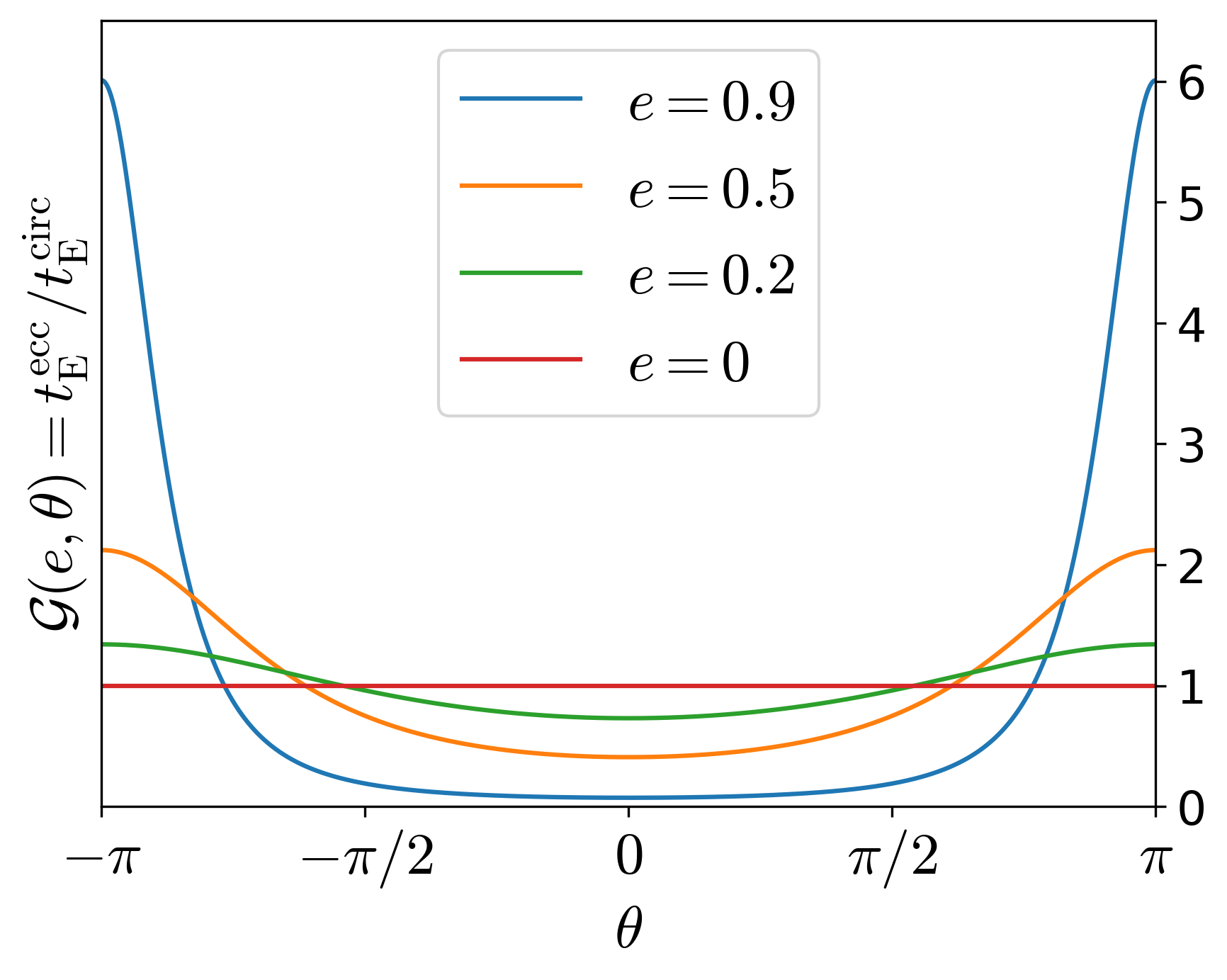}
\caption{Geometrical factor $\mathcal{G}(e,\theta)$ in Eq.~\eqref{eq:te-eccentric}, which corresponds to the difference between the crossing timescale of an eccentric orbit compared to a circular orbit with the same semi-major axis $a$. We plot different values of the orbital eccentricity $e$ compared to the circular case ($e=0$).}
\label{fig:geometrical-factor-eccentric}
\end{figure}

The value of $\te$ of a specific eccentric system is then
\begin{align}
\te (a,e,\theta) &= \frac{2R_{\rm E}}{v_{\rm proj}}
= \frac{4d_{\rm LS}}{c} \frac{1}{\sqrt{1+e\cos\theta}} \nonumber \\
&\equiv \frac{2a}{c}\mathcal{G}(e,\theta)
\label{eq:te-eccentric}
\end{align}
which differs from Eq.~\eqref{eq:tcross} by the geometrical factor 
$\mathcal{G}(e,\theta)=2(1-e^2)(1+e\cos\theta)^{-3/2}$ plotted in Fig.~\ref{fig:geometrical-factor-eccentric}. For $e\lesssim 0.5$, $\te$ lies within the same order of magnitude as the circular case. Highly eccentric orbits, on the other end, offer an opportunity to observe short crossing timescales $\te$ even in orbits with large semi-major axis $a$. However, the probability of 
having the source near to the closest approach (pericenter) is low because the instantaneous velocity is high, and the source spends most of its time far from the pericenter \cite{kepler-1609-first-second-laws,newton-1687-principia}. 

\section{GW chirp waveform}
\label{sec:appendix-chirp}

For this study, we take the inspiral part of the signal\footnote{A CBC signal can be described as having three stages: inspiral (approach of the binary components before merger), merger and ringdown (relaxation into equilibrium of the remnant after the merger).}, because the merger and ringdown are much shorter than $\te$.

The inspiral has a universal evolution ruled by the ``chirp mass'' of the source,  defined as
\begin{equation}
M_{\rm chirp} \equiv \frac{(m_1 m_2)^{3/5}}{(m_1+m_2)^{1/5}}~,
\label{eq:chirp-mass}
\end{equation} 
where $m_1$ and $m_2$ are the intrinsic rest masses of the binary components ($1$ is the most massive and $2$ the least massive). 

\begin{figure*}
\includegraphics[width=0.9\textwidth]{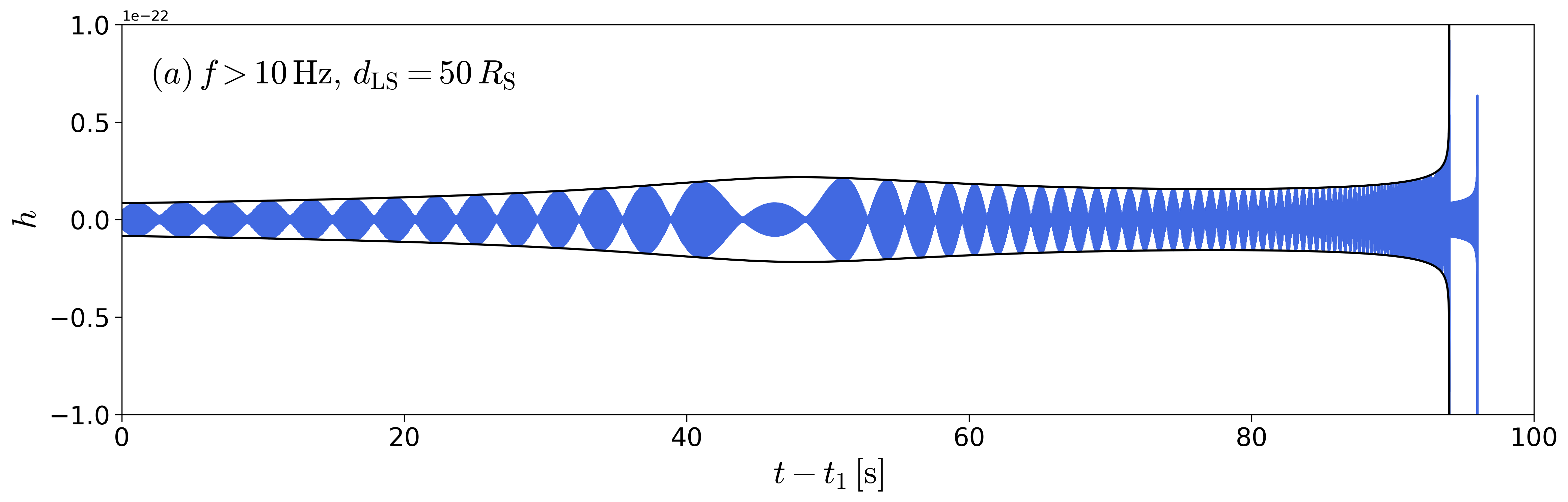}
\includegraphics[width=0.9\textwidth]{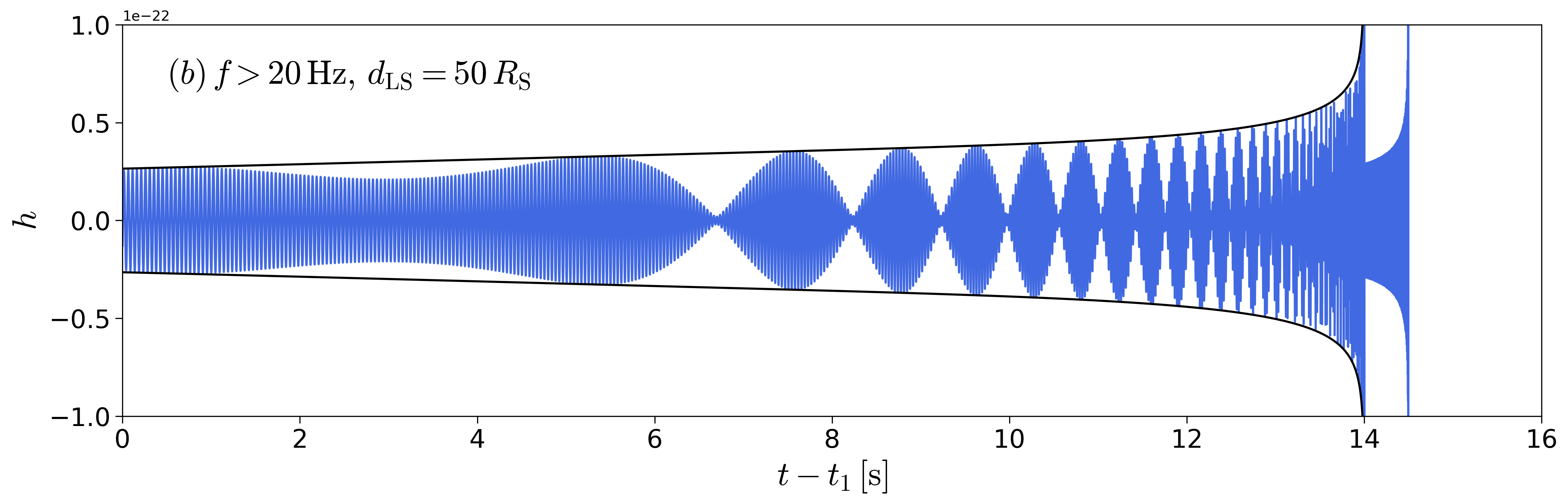}
\includegraphics[width=0.9\textwidth]{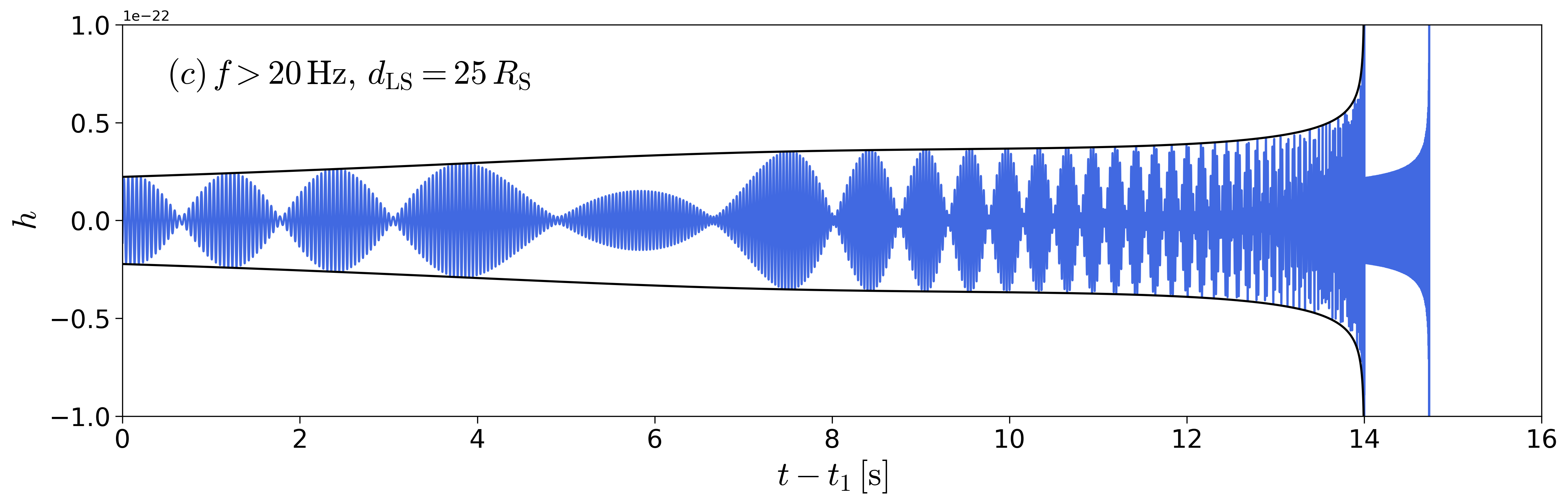}
\caption{Waveform of the same source ($M_{\rm chirp}=5\,M_\odot$, $\mmbhz = 10^5\,M_\odot$) for different lower frequency thresholds $f_{\rm th}$: (a) $f_{\rm th}=10\,{\rm Hz}$; (b,c) $f_{\rm th}=20\,{\rm Hz}$. The crossing time $t_0$ has also been from case to case, fixing it at $t_{\rm insp}/2$ (the shift with respect to the center is because we represent it with respect to $t_1$). The upper plot (a) corresponds to a merger in nominal LVK frequencies $f>10\,{\rm Hz}$. Since the lower frequency is in reality may be affected by data quality, and thus we show the signal for $f>20\,{\rm Hz}$ in comparison (b,c), for two values of the orbital distance in terms of $R_{\rm S}$: at (b) $d_{\rm LS}=50\,R_{\rm S}$; (c) $d_{\rm LS}=25\,R_{\rm S}$.
}
\label{fig:difference-fmin}
\end{figure*}

The waveform for the inspiral of one (unlensed) signal can take any combination of the polarizations $h_+$, $h_\times$ \cite{maggiore-07}:
\begin{equation}
h_+(t) = \frac{1}{d_{\rm S}}
\left(
\frac{G M_{\rm chirp}}{c^2} 
\right)^{5/4} 
\left(
\frac{5}{c \tau} 
\right)^{1/4} 
\left(
\frac{1+\cos^2{\iota}}{2} 
\right)
\cos{[\Phi(\tau)]}~,
\label{eq:hp-ul}
\end{equation}
\begin{equation}
h_\times(t) =
\frac{1}{d_{\rm S}}
\left(
\frac{G M_{\rm chirp}}{c^2} 
\right)^{5/4} 
\left(
\frac{5}{c \tau} 
\right)^{1/4} 
\cos{\iota}\,\,
\sin{[\Phi(\tau)]}~,
\end{equation}
where $\tau = t_{\rm coal}-t$ is the remaining time to the coalescence at $t=t_{\rm coal}$,  and $\iota$ is the inclination of the source ($\iota=0$ for face-on orbital plane, $\iota=1$ for edge-on orbital plane). We consider a BBH merger with edge-on inclination ($\iota=1$) ---due to the self-lensing geometry in the AGN disk---, for which we can use a waveform with pure $h_+$. The phase $\Phi(\tau)$ evolves as
\begin{equation}
\Phi(\tau) = -2 
\left(
\frac{5 G M_{\rm chirp}}{c^3} 
\right)^{-5/8}
\tau^{5/8} + \Phi_0~,
\label{eq:phase-chirp}
\end{equation}
where $\Phi_0=\Phi(\tau=0)$ corresponds to the phase at coalescence.
Because we work in the GO limit, we generate two lensed images evolving as unlensed signals with shifted time variables $\tau_1=t_{\rm coal}-t$ and $\tau_2=t_{\rm coal}-(t+\Delta t)$ (considering them relative to $t_1$). 
The second image includes the Morse phase shift added to the phase $\Phi$. 
These GO images, after applying the magnification (Eq.~\eqref{eq:mu-GO}), can be added up directly in time domain as shown in Eq.~\eqref{eq:h-t-lensed}. 
We use an equal-mass binary (mass ratio $q=1$), with no spin, no precession, and an edge-on viewing orbital inclination. The combination of the images creates the lensing signature shown in Fig.~\ref{fig:lensed_inspiral_waveform}.

\section{Examples of the lensing signature in more limited frequency ranges}
\label{sec:appendix-fmin}

The frequency range of GW detectors (ground-based in this work) will determine if the whole curve is visible. For instance, while for ET the frequency may start at $\sim 3\,{\rm Hz}$, in LVK its starts rather at $\mathcal{O}(10\,{\rm Hz})$. For the same astrophysical parameters ($M_{\rm chirp}=5\,M_\odot$, $\mmbhz = 10^5\,M_\odot$, $d_{\rm LS}=50\,R_{\rm S}$), a difference between starting at $f=10\,{\rm Hz}$ or $f=20\,{\rm Hz}$ significantly reduces the time of the signal, while $\te$ remains the same: in Fig.~\ref{fig:difference-fmin} we show how the evolution of the lensing curve may be more subtle in the $f=20\,{\rm Hz}$ case.

\bibliography{self-lensing-moving-source}

@ARTICLE{paczynski-86,
       author = {{Paczynski}, B.},
        title = "{Gravitational Microlensing by the Galactic Halo}",
      journal = {\apj},
         year = 1986,
        month = may,
       volume = {304},
        pages = {1},
          doi = {10.1086/164140},
       adsurl = {https://ui.adsabs.harvard.edu/abs/1986ApJ...304....1P}
}

@ARTICLE{samsing-25-AGN,
       author = {{Samsing}, Johan and {Zwick}, Lorenz and {Saini}, Pankaj and {Tak{\'a}tsy}, J{\'a}nos},
        title = "{Probing the Formation Environment of Strongly Lensed Black Hole Mergers: Implications for the AGN-disk Channel}",
      journal = {arXiv e-prints},
         year = 2025,
        month = nov,
          eid = {arXiv:2511.19407},
        pages = {arXiv:2511.19407},
          doi = {10.48550/arXiv.2511.19407},
archivePrefix = {arXiv},
       eprint = {2511.19407},
 primaryClass = {astro-ph.HE},
       adsurl = {https://ui.adsabs.harvard.edu/abs/2025arXiv251119407S}
}

@ARTICLE{sudhagar-25,
       author = {{Suyamprakasam}, Sudhagar and {Harikumar}, Sreekanth and {Cieciel{\k{a}}g}, Pawe{\l} and {Figura}, Przemys{\l}aw and {Bejger}, Micha{\l} and {Biesiada}, Marek},
        title = "{Microlensing of long-duration gravitational wave signals originating from Galactic sources}",
      journal = {arXiv e-prints},
         year = 2025,
        month = mar,
          eid = {arXiv:2503.21845},
        pages = {arXiv:2503.21845},
          doi = {10.48550/arXiv.2503.21845},
archivePrefix = {arXiv},
       eprint = {2503.21845},
 primaryClass = {gr-qc},
       adsurl = {https://ui.adsabs.harvard.edu/abs/2025arXiv250321845S}
}

@ARTICLE{li-25,
       author = {{Li}, Yu-Zhe and {Xu}, Wen-Long and {Chen}, Yi-Gu and {Lei}, Wei-Hua},
        title = "{Lense-Thirring Precession Modulates Repeated Lensing of Continues Gravitational Wave Source from AGN Disks}",
      journal = {arXiv e-prints},
         year = 2025,
        month = jun,
          eid = {arXiv:2506.04196},
        pages = {arXiv:2506.04196},
          doi = {10.48550/arXiv.2506.04196},
archivePrefix = {arXiv},
       eprint = {2506.04196},
 primaryClass = {astro-ph.HE},
       adsurl = {https://ui.adsabs.harvard.edu/abs/2025arXiv250604196L}
}

@ARTICLE{santos-25,
       author = {{Santos}, Jo{\~a}o S. and {Cardoso}, Vitor and {Nat{\'a}rio}, Jos{\'e} and {van de Meent}, Maarten},
        title = "{Gravitational Waves from Binary Extreme Mass Ratio Inspirals: Doppler Shift and Beaming, Resonant Excitation, Helicity Oscillations, and Self-Lensing}",
      journal = {\prl},
         year = 2025,
        month = nov,
       volume = {135},
       number = {21},
          eid = {211402},
        pages = {211402},
          doi = {10.1103/qq3m-6phh},
archivePrefix = {arXiv},
       eprint = {2506.14868},
 primaryClass = {gr-qc},
       adsurl = {https://ui.adsabs.harvard.edu/abs/2025PhRvL.135u1402S}
}

@ARTICLE{dorazio-loeb-20,
       author = {{D'Orazio}, Daniel J. and {Loeb}, Abraham},
        title = "{Repeated gravitational lensing of gravitational waves in hierarchical black hole triples}",
      journal = {\prd},
         year = 2020,
        month = apr,
       volume = {101},
       number = {8},
          eid = {083031},
        pages = {083031},
          doi = {10.1103/PhysRevD.101.083031},
archivePrefix = {arXiv},
       eprint = {1910.02966},
 primaryClass = {astro-ph.HE},
       adsurl = {https://ui.adsabs.harvard.edu/abs/2020PhRvD.101h3031D}
}

@ARTICLE{biesiada-harikumar-21,
       author = {{Biesiada}, Marek and {Harikumar}, Sreekanth},
        title = "{Gravitational Lensing of Continuous Gravitational Waves}",
      journal = {Universe},
         year = 2021,
        month = dec,
       volume = {7},
       number = {12},
          eid = {502},
        pages = {502},
          doi = {10.3390/universe7120502},
archivePrefix = {arXiv},
       eprint = {2111.05963},
 primaryClass = {gr-qc},
       adsurl = {https://ui.adsabs.harvard.edu/abs/2021Univ....7..502B}
}

@ARTICLE{liao-19,
       author = {{Liao}, Kai and {Biesiada}, Marek and {Fan}, Xi-Long},
        title = "{The Wave Nature of Continuous Gravitational Waves from Microlensing}",
      journal = {\apj},
         year = 2019,
        month = apr,
       volume = {875},
       number = {2},
          eid = {139},
        pages = {139},
          doi = {10.3847/1538-4357/ab1087},
archivePrefix = {arXiv},
       eprint = {1903.06612},
 primaryClass = {gr-qc},
       adsurl = {https://ui.adsabs.harvard.edu/abs/2019ApJ...875..139L}
}

@ARTICLE{yu-21,
       author = {{Yu}, Hang and {Wang}, Yijun and {Seymour}, Brian and {Chen}, Yanbei},
        title = "{Detecting gravitational lensing in hierarchical triples in galactic nuclei with space-borne gravitational-wave observatories}",
      journal = {\prd},
         year = 2021,
        month = nov,
       volume = {104},
       number = {10},
          eid = {103011},
        pages = {103011},
          doi = {10.1103/PhysRevD.104.103011},
archivePrefix = {arXiv},
       eprint = {2107.14318},
 primaryClass = {gr-qc},
       adsurl = {https://ui.adsabs.harvard.edu/abs/2021PhRvD.104j3011Y}
}

@ARTICLE{ubach-25,
       author = {{Ubach}, Helena and {Gieles}, Mark and {Miralda-Escud{\'e}}, Jordi},
        title = "{Constraining the environment of compact binary mergers with self-lensing signatures}",
      journal = {\prd},
         year = 2025,
        month = oct,
       volume = {112},
       number = {8},
          eid = {083026},
        pages = {083026},
          doi = {10.1103/ql7q-t6wc},
archivePrefix = {arXiv},
       eprint = {2505.04794},
 primaryClass = {astro-ph.HE},
       adsurl = {https://ui.adsabs.harvard.edu/abs/2025PhRvD.112h3026U}
}

@ARTICLE{bulashenko-ubach-22,
       author = {{Bulashenko}, Oleg and {Ubach}, Helena},
        title = "{Lensing of gravitational waves: universal signatures in the beating pattern}",
      journal = {\jcap},
         year = 2022,
        month = jul,
       volume = {2022},
       number = {7},
          eid = {022},
        pages = {022},
          doi = {10.1088/1475-7516/2022/07/022},
archivePrefix = {arXiv},
       eprint = {2112.10773},
 primaryClass = {gr-qc},
       adsurl = {https://ui.adsabs.harvard.edu/abs/2022JCAP...07..022B}
}

@book{maggiore-07,
    author = {Maggiore, Michele},
    title = "{Gravitational Waves: Volume 1: Theory and Experiments}",
    publisher = {Oxford University Press},
    year = {2007},
    month = {10},
    isbn = {9780198570745},
    doi = {10.1093/acprof:oso/9780198570745.001.0001},
    url = {https://doi.org/10.1093/acprof:oso/9780198570745.001.0001},
}

@ARTICLE{depaolis-01,
       author = {{De Paolis}, F. and {Ingrosso}, G. and {Nucita}, A.~A.},
        title = "{Astrophysical implications of gravitational microlensing of gravitational waves}",
      journal = {\aap},
         year = 2001,
        month = feb,
       volume = {366},
        pages = {1065-1070},
          doi = {10.1051/0004-6361:20000304},
archivePrefix = {arXiv},
       eprint = {astro-ph/0011563},
 primaryClass = {astro-ph},
       adsurl = {https://ui.adsabs.harvard.edu/abs/2001A&A...366.1065D}
}

@ARTICLE{hannuksela-19,
       author = {{Hannuksela}, O.~A. and {Haris}, K. and {Ng}, K.~K.~Y. and {Kumar}, S. and {Mehta}, A.~K. and {Keitel}, D. and {Li}, T.~G.~F. and {Ajith}, P.},
        title = "{Search for Gravitational Lensing Signatures in LIGO-Virgo Binary Black Hole Events}",
      journal = {\apjl},
         year = 2019,
        month = mar,
       volume = {874},
       number = {1},
          eid = {L2},
        pages = {L2},
          doi = {10.3847/2041-8213/ab0c0f},
archivePrefix = {arXiv},
       eprint = {1901.02674},
 primaryClass = {gr-qc},
       adsurl = {https://ui.adsabs.harvard.edu/abs/2019ApJ...874L...2H}
}

@Article{LVK-O3a-lensing,
  author        = {{LIGO Scientific Collaboration} and {Virgo Collaboration}},
  journal       = {\apj},
  title         = {{Search for Lensing Signatures in the Gravitational-Wave Observations from the First Half of LIGO-Virgo's Third Observing Run}},
  year          = {2021},
  month         = dec,
  number        = {1},
  pages         = {14},
  volume        = {923},
  adsurl        = {https://ui.adsabs.harvard.edu/abs/2021ApJ...923...14A},
  archiveprefix = {arXiv},
  doi           = {10.3847/1538-4357/ac23db},
  eid           = {14},
  eprint        = {2105.06384},
  primaryclass  = {gr-qc},
}

@ARTICLE{LVK-O3ab-Lensing,
       author = {{The LIGO Scientific Collaboration} and {the Virgo Collaboration} and {the KAGRA Collaboration}},
        title = "{Search for gravitational-lensing signatures in the full third observing run of the LIGO-Virgo network}",
      journal = {arXiv e-prints},
         year = 2023,
        month = apr,
          eid = {arXiv:2304.08393},
        pages = {arXiv:2304.08393},
          doi = {10.48550/arXiv.2304.08393},
archivePrefix = {arXiv},
       eprint = {2304.08393},
 primaryClass = {gr-qc},
       adsurl = {https://ui.adsabs.harvard.edu/abs/2023arXiv230408393T}
}

@ARTICLE{LVK-O4a-Lensing,
       author = {{The LIGO Scientific Collaboration} and {the Virgo Collaboration} and {the KAGRA Collaboration}},
        title = "{GWTC-4.0: Searches for Gravitational-Wave Lensing Signatures}",
      journal = {arXiv e-prints},
         year = 2025,
        month = dec,
          eid = {arXiv:2512.16347},
        pages = {arXiv:2512.16347},
          doi = {10.48550/arXiv.2512.16347},
archivePrefix = {arXiv},
       eprint = {2512.16347},
 primaryClass = {gr-qc},
       adsurl = {https://ui.adsabs.harvard.edu/abs/2025arXiv251216347T}
}

@ARTICLE{goyal-25,
       author = {{Goyal}, Srashti and {Villarrubia-Rojo}, Hector and {Zumalacarregui}, Miguel},
        title = "{Across the Universe: GW231123 as a magnified and diffracted black hole merger}",
      journal = {arXiv e-prints},
         year = 2025,
        month = dec,
          eid = {arXiv:2512.17631},
        pages = {arXiv:2512.17631},
          doi = {10.48550/arXiv.2512.17631},
archivePrefix = {arXiv},
       eprint = {2512.17631},
 primaryClass = {astro-ph.GA},
       adsurl = {https://ui.adsabs.harvard.edu/abs/2025arXiv251217631G}
}

@ARTICLE{chan-25-231123,
       author = {{Chan}, Juno C.~L. and {Mar{\'\i}a Ezquiaga}, Jose and {Lo}, Rico K.~L. and {Bowman}, Joey and {Maga{\~n}a Zertuche}, Lorena and {Vujeva}, Luka},
        title = "{Discovering gravitational waveform distortions from lensing: a deep dive into GW231123}",
      journal = {arXiv e-prints},
         year = 2025,
        month = dec,
          eid = {arXiv:2512.16916},
        pages = {arXiv:2512.16916},
          doi = {10.48550/arXiv.2512.16916},
archivePrefix = {arXiv},
       eprint = {2512.16916},
 primaryClass = {gr-qc},
       adsurl = {https://ui.adsabs.harvard.edu/abs/2025arXiv251216916C}
}

@ARTICLE{hu-25,
       author = {{Hu}, Qian and {Narola}, Harsh and {Heynen}, Jef and {Wright}, Mick and {Veitch}, John and {Janquart}, Justin and {Van Den Broeck}, Chris},
        title = "{GW231123: Overlapping Gravitational Wave Signals?}",
      journal = {arXiv e-prints},
         year = 2025,
        month = dec,
          eid = {arXiv:2512.17550},
        pages = {arXiv:2512.17550},
          doi = {10.48550/arXiv.2512.17550},
archivePrefix = {arXiv},
       eprint = {2512.17550},
 primaryClass = {gr-qc},
       adsurl = {https://ui.adsabs.harvard.edu/abs/2025arXiv251217550H}
}

@ARTICLE{26-chakraborty,
       author = {{Chakraborty}, Aniruddha and {Mukherjee}, Suvodip},
        title = "{The First Model-independent Upper Bound on Microlensing Signature of the Highest-mass Binary Black Hole Event GW231123}",
      journal = {\apj},
         year = 2026,
        month = may,
       volume = {1003},
       number = {1},
          eid = {20},
        pages = {20},
          doi = {10.3847/1538-4357/ae61a7},
archivePrefix = {arXiv},
       eprint = {2512.19077},
 primaryClass = {gr-qc},
       adsurl = {https://ui.adsabs.harvard.edu/abs/2026ApJ..1003...20C}
}

@ARTICLE{25-shan,
       author = {{Shan}, Xikai and {Yang}, Huan and {Mao}, Shude},
        title = "{GW231123: A Case for Binary Microlensing in a Strong Lensing Field}",
      journal = {arXiv e-prints},
         year = 2025,
        month = dec,
          eid = {arXiv:2512.19118},
        pages = {arXiv:2512.19118},
          doi = {10.48550/arXiv.2512.19118},
archivePrefix = {arXiv},
       eprint = {2512.19118},
 primaryClass = {astro-ph.GA},
       adsurl = {https://ui.adsabs.harvard.edu/abs/2025arXiv251219118S}
}

@Article{dai-20,
  author        = {{Dai}, Liang and {Zackay}, Barak and {Venumadhav}, Tejaswi and {Roulet}, Javier and {Zaldarriaga}, Matias},
  journal       = {arXiv e-prints},
  title         = {{Search for Lensed Gravitational Waves Including Morse Phase Information: An Intriguing Candidate in O2}},
  year          = {2020},
  month         = jul,
  pages         = {arXiv:2007.12709},
  adsurl        = {https://ui.adsabs.harvard.edu/abs/2020arXiv200712709D},
  archiveprefix = {arXiv},
  doi           = {10.48550/arXiv.2007.12709},
  eid           = {arXiv:2007.12709},
  eprint        = {2007.12709},
  primaryclass  = {astro-ph.HE},
}

@ARTICLE{janquart-23,
       author = {{Janquart}, J. and {Wright}, M. and {Goyal}, S. and {Chan}, J.~C.~L. and {Ganguly}, A. and {Garr{\'o}n}, {\'A}. and {Keitel}, D. and {Li}, A.~K.~Y. and {Liu}, A. and {Lo}, R.~K.~L. and {Mishra}, A. and {More}, A. and {Phurailatpam}, H. and {Prasia}, P. and {Ajith}, P. and {Biscoveanu}, S. and {Cremonese}, P. and {Cudell}, J.~R. and {Ezquiaga}, J.~M. and {Garcia-Bellido}, J. and {Hannuksela}, O.~A. and {Haris}, K. and {Harry}, I. and {Hendry}, M. and {Husa}, S. and {Kapadia}, S. and {Li}, T.~G.~F. and {Maga{\~n}a Hernandez}, I. and {Mukherjee}, S. and {Seo}, E. and {Van Den Broeck}, C. and {Veitch}, J.},
        title = "{Follow-up analyses to the O3 LIGO-Virgo-KAGRA lensing searches}",
      journal = {\mnras},
         year = 2023,
        month = dec,
       volume = {526},
       number = {3},
        pages = {3832-3860},
          doi = {10.1093/mnras/stad2909},
archivePrefix = {arXiv},
       eprint = {2306.03827},
 primaryClass = {gr-qc},
       adsurl = {https://ui.adsabs.harvard.edu/abs/2023MNRAS.526.3832J}
}

@ARTICLE{231123,
       author = {{Abac}, A.~G. and others},
        title = "{GW231123: A Binary Black Hole Merger with Total Mass 190-265 M$_{{\ensuremath{\odot}}}$}",
      journal = {\apjl},
         year = 2025,
        month = nov,
       volume = {993},
       number = {1},
          eid = {L25},
        pages = {L25},
          doi = {10.3847/2041-8213/ae0c9c},
archivePrefix = {arXiv},
       eprint = {2507.08219},
 primaryClass = {astro-ph.HE},
       adsurl = {https://ui.adsabs.harvard.edu/abs/2025ApJ...993L..25A}
}

@ARTICLE{LIGO18-GWTC1,
       author = {{Abbott}, B.~P. and others},
        title = "{GWTC-1: A Gravitational-Wave Transient Catalog of Compact Binary Mergers Observed by LIGO and Virgo during the First and Second Observing Runs}",
      journal = {Physical Review X},
         year = 2019,
        month = jul,
       volume = {9},
       number = {3},
          eid = {031040},
        pages = {031040},
          doi = {10.1103/PhysRevX.9.031040},
archivePrefix = {arXiv},
       eprint = {1811.12907},
 primaryClass = {astro-ph.HE},
       adsurl = {https://ui.adsabs.harvard.edu/abs/2019PhRvX...9c1040A}
}

@ARTICLE{LIGO20-GWTC2,
       author = {{Abbott}, R. and others},
        title = "{GWTC-2: Compact Binary Coalescences Observed by LIGO and Virgo during the First Half of the Third Observing Run}",
      journal = {Physical Review X},
         year = 2021,
        month = apr,
       volume = {11},
       number = {2},
          eid = {021053},
        pages = {021053},
          doi = {10.1103/PhysRevX.11.021053},
archivePrefix = {arXiv},
       eprint = {2010.14527},
 primaryClass = {gr-qc},
       adsurl = {https://ui.adsabs.harvard.edu/abs/2021PhRvX..11b1053A}
}

@ARTICLE{LIGO21-GWTC3,
       author = {{Abbott}, R. and others},
        title = "{GWTC-3: Compact Binary Coalescences Observed by LIGO and Virgo during the Second Part of the Third Observing Run}",
      journal = {Physical Review X},
         year = 2023,
        month = oct,
       volume = {13},
       number = {4},
          eid = {041039},
        pages = {041039},
          doi = {10.1103/PhysRevX.13.041039},
archivePrefix = {arXiv},
       eprint = {2111.03606},
 primaryClass = {gr-qc},
       adsurl = {https://ui.adsabs.harvard.edu/abs/2023PhRvX..13d1039A}
}

@Article{venumadhav-20,
  author        = {{Venumadhav}, Tejaswi and {Zackay}, Barak and {Roulet}, Javier and {Dai}, Liang and {Zaldarriaga}, Matias},
  journal       = {\prd},
  title         = {{New binary black hole mergers in the second observing run of Advanced LIGO and Advanced Virgo}},
  year          = {2020},
  month         = apr,
  number        = {8},
  pages         = {083030},
  volume        = {101},
  adsurl        = {https://ui.adsabs.harvard.edu/abs/2020PhRvD.101h3030V},
  archiveprefix = {arXiv},
  doi           = {10.1103/PhysRevD.101.083030},
  eid           = {083030},
  eprint        = {1904.07214},
  primaryclass  = {astro-ph.HE},
}

@Article{zackay-21,
  author        = {{Zackay}, Barak and {Dai}, Liang and {Venumadhav}, Tejaswi and {Roulet}, Javier and {Zaldarriaga}, Matias},
  journal       = {\prd},
  title         = {{Detecting gravitational waves with disparate detector responses: Two new binary black hole mergers}},
  year          = {2021},
  month         = sep,
  number        = {6},
  pages         = {063030},
  volume        = {104},
  adsurl        = {https://ui.adsabs.harvard.edu/abs/2021PhRvD.104f3030Z},
  archiveprefix = {arXiv},
  doi           = {10.1103/PhysRevD.104.063030},
  eid           = {063030},
  eprint        = {1910.09528},
  primaryclass  = {astro-ph.HE},
}

@ARTICLE{LIGO25-GWTC4,
       author = {{The LIGO Scientific Collaboration} and {The Virgo Collaboration} and {the KAGRA Collaboration}},
        title = "{GWTC-4.0: Updating the Gravitational-Wave Transient Catalog with Observations from the First Part of the Fourth LIGO-Virgo-KAGRA Observing Run}",
      journal = {arXiv e-prints},
         year = 2025,
        month = aug,
          eid = {arXiv:2508.18082},
        pages = {arXiv:2508.18082},
          doi = {10.48550/arXiv.2508.18082},
archivePrefix = {arXiv},
       eprint = {2508.18082},
 primaryClass = {gr-qc},
       adsurl = {https://ui.adsabs.harvard.edu/abs/2025arXiv250818082T}
}

@Article{LIGO26-GWTC5,
  author        = {{The LIGO Scientific Collaboration} and {the Virgo Collaboration} and {the KAGRA Collaboration}},
  journal       = {arXiv e-prints},
  title         = {{GWTC-5.0: Observations from the Second Part of the Fourth LIGO-Virgo-KAGRA Observing Run and Updates to the Gravitational-Wave Transient Catalog}},
  year          = {2026},
  month         = may,
  pages         = {arXiv:2605.27225},
  adsurl        = {https://ui.adsabs.harvard.edu/abs/2026arXiv260527225T},
  archiveprefix = {arXiv},
  eid           = {arXiv:2605.27225},
  eprint        = {2605.27225},
  primaryclass  = {gr-qc},
}

@Article{wadekar-23,
  author        = {{Wadekar}, Digvijay and {Roulet}, Javier and {Venumadhav}, Tejaswi and {Mehta}, Ajit Kumar and {Zackay}, Barak and {Mushkin}, Jonathan and {Olsen}, Seth and {Zaldarriaga}, Matias},
  journal       = {arXiv e-prints},
  title         = {{New black hole mergers in the LIGO-Virgo O3 data from a gravitational wave search including higher-order harmonics}},
  year          = {2023},
  month         = dec,
  pages         = {arXiv:2312.06631},
  adsurl        = {https://ui.adsabs.harvard.edu/abs/2023arXiv231206631W},
  archiveprefix = {arXiv},
  doi           = {10.48550/arXiv.2312.06631},
  eid           = {arXiv:2312.06631},
  eprint        = {2312.06631},
  primaryclass  = {gr-qc},
}

@ARTICLE{ng-18,
       author = {{Ng}, Ken K.~Y. and {Wong}, Kaze W.~K. and {Broadhurst}, Tom and {Li}, Tjonnie G.~F.},
        title = "{Precise LIGO lensing rate predictions for binary black holes}",
      journal = {\prd},
         year = 2018,
        month = jan,
       volume = {97},
       number = {2},
          eid = {023012},
        pages = {023012},
          doi = {10.1103/PhysRevD.97.023012},
archivePrefix = {arXiv},
       eprint = {1703.06319},
 primaryClass = {astro-ph.CO},
       adsurl = {https://ui.adsabs.harvard.edu/abs/2018PhRvD..97b3012N}
}

@ARTICLE{li-18,
       author = {{Li}, Shun-Sheng and {Mao}, Shude and {Zhao}, Yuetong and {Lu}, Youjun},
        title = "{Gravitational lensing of gravitational waves: a statistical perspective}",
      journal = {\mnras},
         year = 2018,
        month = may,
       volume = {476},
       number = {2},
        pages = {2220-2229},
          doi = {10.1093/mnras/sty411},
archivePrefix = {arXiv},
       eprint = {1802.05089},
 primaryClass = {astro-ph.CO},
       adsurl = {https://ui.adsabs.harvard.edu/abs/2018MNRAS.476.2220L}
}

@ARTICLE{oguri-18,
       author = {{Oguri}, Masamune},
        title = "{Effect of gravitational lensing on the distribution of gravitational waves from distant binary black hole mergers}",
      journal = {\mnras},
         year = 2018,
        month = nov,
       volume = {480},
       number = {3},
        pages = {3842-3855},
          doi = {10.1093/mnras/sty2145},
archivePrefix = {arXiv},
       eprint = {1807.02584},
 primaryClass = {astro-ph.CO},
       adsurl = {https://ui.adsabs.harvard.edu/abs/2018MNRAS.480.3842O}
}

@ARTICLE{mukherjee-21,
       author = {{Mukherjee}, Suvodip and {Broadhurst}, Tom and {Diego}, Jose M. and {Silk}, Joseph and {Smoot}, George F.},
        title = "{Impact of astrophysical binary coalescence time-scales on the rate of lensed gravitational wave events}",
      journal = {\mnras},
         year = 2021,
        month = sep,
       volume = {506},
       number = {3},
        pages = {3751-3759},
          doi = {10.1093/mnras/stab1980},
archivePrefix = {arXiv},
       eprint = {2106.00392},
 primaryClass = {gr-qc},
       adsurl = {https://ui.adsabs.harvard.edu/abs/2021MNRAS.506.3751M}
}

@ARTICLE{wierda-21,
       author = {{Wierda}, A. Renske A.~C. and {Wempe}, Ewoud and {Hannuksela}, Otto A. and {Koopmans}, L{\'e}on V.~E. and {Van Den Broeck}, Chris},
        title = "{Beyond the Detector Horizon: Forecasting Gravitational-Wave Strong Lensing}",
      journal = {\apj},
         year = 2021,
        month = nov,
       volume = {921},
       number = {2},
          eid = {154},
        pages = {154},
          doi = {10.3847/1538-4357/ac1bb4},
archivePrefix = {arXiv},
       eprint = {2106.06303},
 primaryClass = {astro-ph.HE},
       adsurl = {https://ui.adsabs.harvard.edu/abs/2021ApJ...921..154W}
}

@ARTICLE{xu-ezquiaga-holz-22,
       author = {{Xu}, Fei and {Ezquiaga}, Jose Mar{\'\i}a and {Holz}, Daniel E.},
        title = "{Please Repeat: Strong Lensing of Gravitational Waves as a Probe of Compact Binary and Galaxy Populations}",
      journal = {\apj},
         year = 2022,
        month = apr,
       volume = {929},
       number = {1},
          eid = {9},
        pages = {9},
          doi = {10.3847/1538-4357/ac58f8},
archivePrefix = {arXiv},
       eprint = {2105.14390},
 primaryClass = {astro-ph.CO},
       adsurl = {https://ui.adsabs.harvard.edu/abs/2022ApJ...929....9X}
}

@ARTICLE{walsh-79,
       author = {{Walsh}, D. and {Carswell}, R.~F. and {Weymann}, R.~J.},
        title = "{0957+561 A, B: twin quasistellar objects or gravitational lens?}",
      journal = {\nat},
         year = 1979,
        month = may,
       volume = {279},
        pages = {381-384},
          doi = {10.1038/279381a0},
       adsurl = {https://ui.adsabs.harvard.edu/abs/1979Natur.279..381W}
}

@ARTICLE{alcock-93,
       author = {{Alcock}, C. and {Akerlof}, C.~W. and {Allsman}, R.~A. and {Axelrod}, T.~S. and {Bennett}, D.~P. and {Chan}, S. and {Cook}, K.~H. and {Freeman}, K.~C. and {Griest}, K. and {Marshall}, S.~L. and {Park}, H.-S. and {Perlmutter}, S. and {Peterson}, B.~A. and {Pratt}, M.~R. and {Quinn}, P.~J. and {Rodgers}, A.~W. and {Stubbs}, C.~W. and {Sutherland}, W.},
        title = "{Possible gravitational microlensing of a star in the Large Magellanic Cloud}",
      journal = {\nat},
         year = 1993,
        month = oct,
       volume = {365},
       number = {6447},
        pages = {621-623},
          doi = {10.1038/365621a0},
archivePrefix = {arXiv},
       eprint = {astro-ph/9309052},
 primaryClass = {astro-ph},
       adsurl = {https://ui.adsabs.harvard.edu/abs/1993Natur.365..621A}
}

@ARTICLE{udalski-93,
       author = {{Udalski}, A. and {Szymanski}, M. and {Kaluzny}, J. and {Kubiak}, M. and {Krzeminski}, W. and {Mateo}, M. and {Preston}, G.~W. and {Paczynski}, B.},
        title = "{The Optical Gravitational Lensing Experiment. Discovery of the First Candidate Microlensing Event in the Direction of the Galactic Bulge}",
      journal = {\actaa},
         year = 1993,
        month = jul,
       volume = {43},
        pages = {289-294},
       adsurl = {https://ui.adsabs.harvard.edu/abs/1993AcA....43..289U}
}

@ARTICLE{gould-00,
       author = {{Gould}, Andrew},
        title = "{A Natural Formalism for Microlensing}",
      journal = {\apj},
         year = 2000,
        month = oct,
       volume = {542},
       number = {2},
        pages = {785-788},
          doi = {10.1086/317037},
archivePrefix = {arXiv},
       eprint = {astro-ph/0001421},
 primaryClass = {astro-ph},
       adsurl = {https://ui.adsabs.harvard.edu/abs/2000ApJ...542..785G}
}

@ARTICLE{gould-19,
       author = {{Gould}, Andrew},
        title = "{Osculating Versus Intersecting Circles in Space-Based Microlens Parallax Degeneracies}",
      journal = {Journal of Korean Astronomical Society},
         year = 2019,
        month = aug,
       volume = {52},
       number = {4},
        pages = {121-131},
          doi = {10.5303/JKAS.2019.52.4.121},
archivePrefix = {arXiv},
       eprint = {1905.06770},
 primaryClass = {astro-ph.EP},
       adsurl = {https://ui.adsabs.harvard.edu/abs/2019JKAS...52..121G}
}

@ARTICLE{rahvar-20,
       author = {{Rahvar}, Sohrab},
        title = "{Frequency-shift in the gravitational microlensing}",
      journal = {\prd},
         year = 2020,
        month = jan,
       volume = {101},
       number = {2},
          eid = {024015},
        pages = {024015},
          doi = {10.1103/PhysRevD.101.024015},
archivePrefix = {arXiv},
       eprint = {1908.01361},
 primaryClass = {astro-ph.GA},
       adsurl = {https://ui.adsabs.harvard.edu/abs/2020PhRvD.101b4015R}
}

@ARTICLE{lee-17,
       author = {{Lee}, Chien-Hsiu},
        title = "{Microlensing and Its Degeneracy Breakers: Parallax, Finite Source, High-Resolution Imaging, and Astrometry}",
      journal = {Universe},
         year = 2017,
        month = jul,
       volume = {3},
       number = {3},
          eid = {53},
        pages = {53},
          doi = {10.3390/universe3030053},
archivePrefix = {arXiv},
       eprint = {1711.05298},
 primaryClass = {astro-ph.IM},
       adsurl = {https://ui.adsabs.harvard.edu/abs/2017Univ....3...53L}
}

@ARTICLE{gould-94a,
       author = {{Gould}, Andrew},
        title = "{Proper Motions of MACHOs}",
      journal = {\apjl},
         year = 1994,
        month = feb,
       volume = {421},
        pages = {L71},
          doi = {10.1086/187190},
       adsurl = {https://ui.adsabs.harvard.edu/abs/1994ApJ...421L..71G}
}

@ARTICLE{gould-94b,
       author = {{Gould}, Andrew},
        title = "{MACHO Velocities from Satellite-based Parallaxes}",
      journal = {\apjl},
         year = 1994,
        month = feb,
       volume = {421},
        pages = {L75},
          doi = {10.1086/187191},
       adsurl = {https://ui.adsabs.harvard.edu/abs/1994ApJ...421L..75G}
}

@ARTICLE{nemiroff-94,
       author = {{Nemiroff}, Robert J. and {Wickramasinghe}, W.~A.~D.~T.},
        title = "{Finite Source Sizes and the Information Content of Macho-Type Lens Search Light Curves}",
      journal = {\apjl},
         year = 1994,
        month = mar,
       volume = {424},
        pages = {L21},
          doi = {10.1086/187265},
archivePrefix = {arXiv},
       eprint = {astro-ph/9401005},
 primaryClass = {astro-ph},
       adsurl = {https://ui.adsabs.harvard.edu/abs/1994ApJ...424L..21N}
}

@ARTICLE{refsdal-66,
       author = {{Refsdal}, S.},
        title = "{On the Possibility of Determining the Distances and Masses of Stars from the Gravitational Lens Effect}",
      journal = {\mnras},
         year = 1966,
        month = dec,
       volume = {134},
       number = {3},
        pages = {315-319},
          doi = {10.1093/mnras/134.3.315},
       adsurl = {https://ui.adsabs.harvard.edu/abs/1966MNRAS.134..315R}
}

@ARTICLE{beskin-tuntsov-02,
       author = {{Beskin}, G.~M. and {Tuntsov}, A.~V.},
        title = "{Detection of compact objects by means of gravitational lensing in binary systems}",
      journal = {\aap},
         year = 2002,
        month = nov,
       volume = {394},
        pages = {489-503},
          doi = {10.1051/0004-6361:20021150},
archivePrefix = {arXiv},
       eprint = {astro-ph/0208095},
 primaryClass = {astro-ph},
       adsurl = {https://ui.adsabs.harvard.edu/abs/2002A&A...394..489B}
}

@ARTICLE{rahvar-11,
       author = {{Rahvar}, S. and {Mehrabi}, A. and {Dominik}, M.},
        title = "{Compact object detection in self-lensing binary systems with a main-sequence star}",
      journal = {\mnras},
         year = 2011,
        month = jan,
       volume = {410},
       number = {2},
        pages = {912-918},
          doi = {10.1111/j.1365-2966.2010.17490.x},
archivePrefix = {arXiv},
       eprint = {1008.1033},
 primaryClass = {astro-ph.SR},
       adsurl = {https://ui.adsabs.harvard.edu/abs/2011MNRAS.410..912R}
}

@ARTICLE{kasuya-11,
       author = {{Kasuya}, Shinta and {Honda}, Mitsuhiko and {Mishima}, Risa},
        title = "{New observable for gravitational lensing effects during transits}",
      journal = {\mnras},
         year = 2011,
        month = mar,
       volume = {411},
       number = {3},
        pages = {1863-1868},
          doi = {10.1111/j.1365-2966.2010.17809.x},
archivePrefix = {arXiv},
       eprint = {1009.2129},
 primaryClass = {astro-ph.EP},
       adsurl = {https://ui.adsabs.harvard.edu/abs/2011MNRAS.411.1863K}
}

@ARTICLE{han-16,
       author = {{Han}, Cheongho},
        title = "{Degeneracy between Lensing and Occultation in the Analysis of Self-lensing Phenomena}",
      journal = {\apj},
         year = 2016,
        month = mar,
       volume = {820},
       number = {1},
          eid = {53},
        pages = {53},
          doi = {10.3847/0004-637X/820/1/53},
archivePrefix = {arXiv},
       eprint = {1603.03500},
 primaryClass = {astro-ph.SR},
       adsurl = {https://ui.adsabs.harvard.edu/abs/2016ApJ...820...53H}
}

@ARTICLE{savastano-24,
       author = {{Savastano}, Stefano and {Vernizzi}, Filippo and {Zumalac{\'a}rregui}, Miguel},
        title = "{Through the lens of Sgr A$^{*}$: Identifying and resolving strongly lensed continuous gravitational waves beyond the Einstein radius}",
      journal = {\prd},
         year = 2024,
        month = jan,
       volume = {109},
       number = {2},
          eid = {024064},
        pages = {024064},
          doi = {10.1103/PhysRevD.109.024064},
archivePrefix = {arXiv},
       eprint = {2212.14697},
 primaryClass = {gr-qc},
       adsurl = {https://ui.adsabs.harvard.edu/abs/2024PhRvD.109b4064S}
}

@ARTICLE{yang-25,
       author = {{Yang}, Xing-Yu and {Chen}, Tan and {Cai}, Rong-Gen},
        title = "{Modulations of gravitational waves due to non-static gravitational lenses}",
      journal = {\jcap},
         year = 2025,
        month = apr,
       volume = {2025},
       number = {4},
          eid = {069},
        pages = {069},
          doi = {10.1088/1475-7516/2025/04/069},
archivePrefix = {arXiv},
       eprint = {2410.16378},
 primaryClass = {astro-ph.CO},
       adsurl = {https://ui.adsabs.harvard.edu/abs/2025JCAP...04..069Y}
}

@ARTICLE{guo-24,
       author = {{Guo}, Xiao and {Cao}, Zhoujian},
        title = "{Testing an exact diffraction formula with gravitational wave source lensed by a supermassive black hole in binary systems}",
      journal = {\jcap},
         year = 2024,
        month = may,
       volume = {2024},
       number = {5},
          eid = {084},
        pages = {084},
          doi = {10.1088/1475-7516/2024/05/084},
archivePrefix = {arXiv},
       eprint = {2401.01581},
 primaryClass = {astro-ph.HE},
       adsurl = {https://ui.adsabs.harvard.edu/abs/2024JCAP...05..084G}
}

@ARTICLE{hou-20,
       author = {{Hou}, Shaoqi and {Fan}, Xi-Long and {Liao}, Kai and {Zhu}, Zong-Hong},
        title = "{Gravitational wave interference via gravitational lensing: Measurements of luminosity distance, lens mass, and cosmological parameters}",
      journal = {\prd},
         year = 2020,
        month = mar,
       volume = {101},
       number = {6},
          eid = {064011},
        pages = {064011},
          doi = {10.1103/PhysRevD.101.064011},
archivePrefix = {arXiv},
       eprint = {1911.02798},
 primaryClass = {gr-qc},
       adsurl = {https://ui.adsabs.harvard.edu/abs/2020PhRvD.101f4011H}
}

@ARTICLE{sun-fan-19,
       author = {{Sun}, Dongze and {Fan}, Xilong},
        title = "{Pattern of lensed chirp gravitational wave signal and its implication on the mass and position of lens}",
      journal = {arXiv e-prints},
         year = 2019,
        month = nov,
          eid = {arXiv:1911.08268},
        pages = {arXiv:1911.08268},
          doi = {10.48550/arXiv.1911.08268},
archivePrefix = {arXiv},
       eprint = {1911.08268},
 primaryClass = {gr-qc},
       adsurl = {https://ui.adsabs.harvard.edu/abs/2019arXiv191108268S}
}

@ARTICLE{samsing-25-GWpulsar,
       author = {{Samsing}, Johan and {Hendriks}, Kai and {Zwick}, Lorenz and {D'Orazio}, Daniel J. and {Liu}, Bin},
        title = "{Gravitational-wave Phase Shifts in Eccentric Black Hole Mergers as a Probe of Dynamical Formation Environments}",
      journal = {\apj},
         year = 2025,
        month = sep,
       volume = {990},
       number = {2},
          eid = {211},
        pages = {211},
          doi = {10.3847/1538-4357/ad9f3d},
archivePrefix = {arXiv},
       eprint = {2403.05625},
 primaryClass = {astro-ph.HE},
       adsurl = {https://ui.adsabs.harvard.edu/abs/2025ApJ...990..211S}
}

@ARTICLE{gondan-kocsis-22,
       author = {{Gond{\'a}n}, L{\'a}szl{\'o} and {Kocsis}, Bence},
        title = "{Astrophysical gravitational-wave echoes from galactic nuclei}",
      journal = {\mnras},
         year = 2022,
        month = sep,
       volume = {515},
       number = {3},
        pages = {3299-3318},
          doi = {10.1093/mnras/stac1985},
archivePrefix = {arXiv},
       eprint = {2110.09540},
 primaryClass = {astro-ph.HE},
       adsurl = {https://ui.adsabs.harvard.edu/abs/2022MNRAS.515.3299G}
}

@Article{leong-25,
  author        = {{Leong}, Samson H.~W. and {Janquart}, Justin and {Sharma}, Aditya Kumar and {Martens}, Paul and {Ajith}, Parameswaran and {Hannuksela}, Otto A.},
  journal       = {\apjl},
  title         = {{Constraining Binary Mergers in Active Galactic Nuclei Disks Using the Nonobservation of Lensed Gravitational Waves}},
  year          = {2025},
  month         = feb,
  number        = {2},
  pages         = {L27},
  volume        = {979},
  adsurl        = {https://ui.adsabs.harvard.edu/abs/2025ApJ...979L..27L},
  archiveprefix = {arXiv},
  doi           = {10.3847/2041-8213/ad9ead},
  eid           = {L27},
  eprint        = {2408.13144},
  primaryclass  = {astro-ph.HE},
}

@ARTICLE{takatsy-25,
       author = {{Tak{\'a}tsy}, J{\'a}nos and {Zwick}, Lorenz and {Hendriks}, Kai and {Saini}, Pankaj and {Fabj}, Gaia and {Samsing}, Johan},
        title = "{The construction and use of dephasing prescriptions for environmental effects in gravitational wave astronomy}",
      journal = {Classical and Quantum Gravity},
         year = 2025,
        month = nov,
       volume = {42},
       number = {21},
          eid = {215006},
        pages = {215006},
          doi = {10.1088/1361-6382/ae0fd4},
archivePrefix = {arXiv},
       eprint = {2505.09513},
 primaryClass = {astro-ph.HE},
       adsurl = {https://ui.adsabs.harvard.edu/abs/2025CQGra..42u5006T}
}

@ARTICLE{michalowski-21,
       author = {{Micha{\l}owski}, Micha{\l} J. and {Mr{\'o}z}, Przemek},
        title = "{Stars Lensed by the Supermassive Black Hole in the Center of the Milky Way: Predictions for ELT, TMT, GMT, and JWST}",
      journal = {\apjl},
         year = 2021,
        month = jul,
       volume = {915},
       number = {2},
          eid = {L33},
        pages = {L33},
          doi = {10.3847/2041-8213/ac0f81},
archivePrefix = {arXiv},
       eprint = {2107.00659},
 primaryClass = {astro-ph.GA},
       adsurl = {https://ui.adsabs.harvard.edu/abs/2021ApJ...915L..33M}
}

@ARTICLE{bozza-mancini-12,
       author = {{Bozza}, V. and {Mancini}, L.},
        title = "{Observing Gravitational Lensing Effects by Sgr A* with GRAVITY}",
      journal = {\apj},
         year = 2012,
        month = jul,
       volume = {753},
       number = {1},
          eid = {56},
        pages = {56},
          doi = {10.1088/0004-637X/753/1/56},
archivePrefix = {arXiv},
       eprint = {1204.2103},
 primaryClass = {astro-ph.GA},
       adsurl = {https://ui.adsabs.harvard.edu/abs/2012ApJ...753...56B}
}

@ARTICLE{takahashi-03,
       author = {{Takahashi}, Ryuichi and {Nakamura}, Takashi},
        title = "{Wave Effects in the Gravitational Lensing of Gravitational Waves from Chirping Binaries}",
      journal = {\apj},
         year = 2003,
        month = oct,
       volume = {595},
       number = {2},
        pages = {1039-1051},
          doi = {10.1086/377430},
archivePrefix = {arXiv},
       eprint = {astro-ph/0305055},
 primaryClass = {astro-ph},
       adsurl = {https://ui.adsabs.harvard.edu/abs/2003ApJ...595.1039T}
}

@ARTICLE{bellovary-16,
       author = {{Bellovary}, Jillian M. and {Mac Low}, Mordecai-Mark and {McKernan}, Barry and {Ford}, K.~E. Saavik},
        title = "{Migration Traps in Disks around Supermassive Black Holes}",
      journal = {\apjl},
         year = 2016,
        month = mar,
       volume = {819},
       number = {2},
          eid = {L17},
        pages = {L17},
          doi = {10.3847/2041-8205/819/2/L17},
archivePrefix = {arXiv},
       eprint = {1511.00005},
 primaryClass = {astro-ph.GA},
       adsurl = {https://ui.adsabs.harvard.edu/abs/2016ApJ...819L..17B}
}

@ARTICLE{secunda-19,
       author = {{Secunda}, Amy and {Bellovary}, Jillian and {Mac Low}, Mordecai-Mark and {Ford}, K.~E. Saavik and {McKernan}, Barry and {Leigh}, Nathan W.~C. and {Lyra}, Wladimir and {S{\'a}ndor}, Zsolt},
        title = "{Orbital Migration of Interacting Stellar Mass Black Holes in Disks around Supermassive Black Holes}",
      journal = {\apj},
         year = 2019,
        month = jun,
       volume = {878},
       number = {2},
          eid = {85},
        pages = {85},
          doi = {10.3847/1538-4357/ab20ca},
       adsurl = {https://ui.adsabs.harvard.edu/abs/2019ApJ...878...85S}
}

@Article{grishin-24,
  author        = {{Grishin}, Evgeni and {Gilbaum}, Shmuel and {Stone}, Nicholas C.},
  journal       = {\mnras},
  title         = {{The effect of thermal torques on AGN disc migration traps and gravitational wave populations}},
  year          = {2024},
  month         = may,
  number        = {2},
  pages         = {2114-2132},
  volume        = {530},
  adsurl        = {https://ui.adsabs.harvard.edu/abs/2024MNRAS.530.2114G},
  archiveprefix = {arXiv},
  doi           = {10.1093/mnras/stae828},
  eprint        = {2307.07546},
  primaryclass  = {astro-ph.HE},
}

@ARTICLE{peng-21,
       author = {{Peng}, Peng and {Chen}, Xian},
        title = "{The last migration trap of compact objects in AGN accretion disc}",
      journal = {\mnras},
         year = 2021,
        month = jul,
       volume = {505},
       number = {1},
        pages = {1324-1333},
          doi = {10.1093/mnras/stab1419},
archivePrefix = {arXiv},
       eprint = {2104.07685},
 primaryClass = {astro-ph.HE},
       adsurl = {https://ui.adsabs.harvard.edu/abs/2021MNRAS.505.1324P}
}

@ARTICLE{mckernan-12,
       author = {{McKernan}, B. and {Ford}, K.~E.~S. and {Lyra}, W. and {Perets}, H.~B.},
        title = "{Intermediate mass black holes in AGN discs - I. Production and growth}",
      journal = {\mnras},
         year = 2012,
        month = sep,
       volume = {425},
       number = {1},
        pages = {460-469},
          doi = {10.1111/j.1365-2966.2012.21486.x},
archivePrefix = {arXiv},
       eprint = {1206.2309},
 primaryClass = {astro-ph.GA},
       adsurl = {https://ui.adsabs.harvard.edu/abs/2012MNRAS.425..460M}
}

@ARTICLE{gangardt-24,
       author = {{Gangardt}, Daria and {Trani}, Alessandro Alberto and {Bonnerot}, Cl{\'e}ment and {Gerosa}, Davide},
        title = "{pAGN: the one-stop solution for AGN disc modelling}",
      journal = {\mnras},
         year = 2024,
        month = jun,
       volume = {530},
       number = {4},
        pages = {3689-3705},
          doi = {10.1093/mnras/stae1117},
archivePrefix = {arXiv},
       eprint = {2403.00060},
 primaryClass = {astro-ph.HE},
       adsurl = {https://ui.adsabs.harvard.edu/abs/2024MNRAS.530.3689G}
}

@ARTICLE{rowan-25a,
       author = {{Rowan}, Connar and {Whitehead}, Henry and {Fabj}, Gaia and {Kirkeberg}, Philip and {Pessah}, Martin E. and {Kocsis}, Bence},
        title = "{Hydrodynamic simulations of black hole evolution in AGN discs {\textendash} I. Orbital alignment of highly inclined satellites}",
      journal = {\mnras},
         year = 2025,
        month = oct,
       volume = {543},
       number = {1},
        pages = {132-145},
          doi = {10.1093/mnras/staf1449},
archivePrefix = {arXiv},
       eprint = {2505.23739},
 primaryClass = {astro-ph.HE},
       adsurl = {https://ui.adsabs.harvard.edu/abs/2025MNRAS.543..132R}
}

@ARTICLE{nakamura-98,
       author = {{Nakamura}, Takahiro T.},
        title = "{Gravitational Lensing of Gravitational Waves from Inspiraling Binaries by a Point Mass Lens}",
      journal = {\prl},
         year = 1998,
        month = feb,
       volume = {80},
       number = {6},
        pages = {1138-1141},
          doi = {10.1103/PhysRevLett.80.1138},
       adsurl = {https://ui.adsabs.harvard.edu/abs/1998PhRvL..80.1138N}
}

@ARTICLE{zwicky-37,
       author = {{Zwicky}, F.},
        title = "{Nebulae as Gravitational Lenses}",
      journal = {Physical Review},
         year = 1937,
        month = feb,
       volume = {51},
       number = {4},
        pages = {290-290},
          doi = {10.1103/PhysRev.51.290},
       adsurl = {https://ui.adsabs.harvard.edu/abs/1937PhRv...51..290Z}
}

@ARTICLE{paczynski-95,
       author = {{Paczynski}, B.},
        title = "{The Masses of Nearby Dwarfs Can Be Determined with Gravitational Microlensing}",
      journal = {\actaa},
         year = 1995,
        month = apr,
       volume = {45},
        pages = {345-348},
          doi = {10.48550/arXiv.astro-ph/9504099},
archivePrefix = {arXiv},
       eprint = {astro-ph/9504099},
 primaryClass = {astro-ph},
       adsurl = {https://ui.adsabs.harvard.edu/abs/1995AcA....45..345P}
}

@ARTICLE{wang-96,
       author = {{Wang}, Yun and {Stebbins}, Albert and {Turner}, Edwin L.},
        title = "{Gravitational Lensing of Gravitational Waves from Merging Neutron Star Binaries}",
      journal = {\prl},
         year = 1996,
        month = sep,
       volume = {77},
       number = {14},
        pages = {2875-2878},
          doi = {10.1103/PhysRevLett.77.2875},
archivePrefix = {arXiv},
       eprint = {astro-ph/9605140},
 primaryClass = {astro-ph},
       adsurl = {https://ui.adsabs.harvard.edu/abs/1996PhRvL..77.2875W}
}

@ARTICLE{narayan-bartelmann-96,
       author = {{Narayan}, Ramesh and {Bartelmann}, Matthias},
        title = "{Lectures on Gravitational Lensing}",
      journal = {arXiv e-prints},
         year = 1996,
        month = jun,
          eid = {astro-ph/9606001},
        pages = {astro-ph/9606001},
          doi = {10.48550/arXiv.astro-ph/9606001},
archivePrefix = {arXiv},
       eprint = {astro-ph/9606001},
 primaryClass = {astro-ph},
       adsurl = {https://ui.adsabs.harvard.edu/abs/1996astro.ph..6001N}
}

@ARTICLE{trani-di-cintio-25,
       author = {{Trani}, Alessandro Alberto and {Di Cintio}, Pierfrancesco},
        title = "{Turbulent drag on stellar mass black holes embedded in disks of active galactic nuclei}",
      journal = {\aap},
         year = 2025,
        month = oct,
       volume = {703},
          eid = {A6},
        pages = {A6},
          doi = {10.1051/0004-6361/202555879},
archivePrefix = {arXiv},
       eprint = {2506.02173},
 primaryClass = {astro-ph.HE},
       adsurl = {https://ui.adsabs.harvard.edu/abs/2025A&A...703A...6T}
}

@ARTICLE{postiglione-25,
       author = {{Postiglione}, Jake and {Ford}, K.~E. Saavik and {Best}, Henry and {McKernan}, Barry and {O'Dowd}, Matthew},
        title = "{Evolution of LISA Observables for Binary Black Holes Lensed by a Supermassive Black Hole}",
      journal = {\apj},
         year = 2025,
        month = oct,
       volume = {991},
       number = {2},
          eid = {161},
        pages = {161},
          doi = {10.3847/1538-4357/adfa0c},
archivePrefix = {arXiv},
       eprint = {2502.10591},
 primaryClass = {astro-ph.HE},
       adsurl = {https://ui.adsabs.harvard.edu/abs/2025ApJ...991..161P}
}

@Article{jow-20,
  author        = {{Jow}, Dylan L. and {Foreman}, Simon and {Pen}, Ue-Li and {Zhu}, Wei},
  journal       = {\mnras},
  title         = {{Wave effects in the microlensing of pulsars and FRBs by point masses}},
  year          = {2020},
  month         = oct,
  number        = {4},
  pages         = {4956-4969},
  volume        = {497},
  adsurl        = {https://ui.adsabs.harvard.edu/abs/2020MNRAS.497.4956J},
  archiveprefix = {arXiv},
  doi           = {10.1093/mnras/staa2230},
  eprint        = {2002.01570},
  primaryclass  = {astro-ph.HE},
}

@Article{itoh-09,
  author        = {{Itoh}, Yousuke and {Futamase}, Toshifumi and {Hattori}, Makoto},
  journal       = {\prd},
  title         = {{Method to measure a relative transverse velocity of a source-lens-observer system using gravitational lensing of gravitational waves}},
  year          = {2009},
  month         = aug,
  number        = {4},
  pages         = {044009},
  volume        = {80},
  adsurl        = {https://ui.adsabs.harvard.edu/abs/2009PhRvD..80d4009I},
  archiveprefix = {arXiv},
  doi           = {10.1103/PhysRevD.80.044009},
  eid           = {044009},
  eprint        = {0908.0186},
  primaryclass  = {gr-qc},
}

@Article{basak-23,
  author        = {{Basak}, Soummyadip and {Sharma}, Aditya Kumar and {Kapadia}, Shasvath J. and {Ajith}, Parameswaran},
  journal       = {\apjl},
  title         = {{Prospects for the Observation of Continuous Gravitational Waves from Spinning Neutron Stars Lensed by the Galactic Supermassive Black Hole}},
  year          = {2023},
  month         = jan,
  number        = {2},
  pages         = {L31},
  volume        = {942},
  adsurl        = {https://ui.adsabs.harvard.edu/abs/2023ApJ...942L..31B},
  archiveprefix = {arXiv},
  doi           = {10.3847/2041-8213/acab50},
  eid           = {L31},
  eprint        = {2205.00022},
  primaryclass  = {gr-qc},
}

@Article{dorazio-distefano-20,
  author        = {{D'Orazio}, Daniel J. and {Di Stefano}, Rosanne},
  journal       = {\mnras},
  title         = {{Detecting gravitational self-lensing from stellar-mass binaries composed of black holes or neutron stars}},
  year          = {2020},
  month         = jan,
  number        = {1},
  pages         = {1506-1517},
  volume        = {491},
  adsurl        = {https://ui.adsabs.harvard.edu/abs/2020MNRAS.491.1506D},
  archiveprefix = {arXiv},
  doi           = {10.1093/mnras/stz3086},
  eprint        = {1906.11149},
  primaryclass  = {astro-ph.HE},
}

@Book{kippenhahn-weigert-90,
  author  = {{Kippenhahn}, Rudolf and {Weigert}, Alfred},
  title   = {{Stellar Structure and Evolution}},
  year    = {1990},
  adsurl  = {https://ui.adsabs.harvard.edu/abs/1990sse..book.....K},
}

@Article{chen-kipping-17,
  author        = {{Chen}, Jingjing and {Kipping}, David},
  journal       = {\apj},
  title         = {{Probabilistic Forecasting of the Masses and Radii of Other Worlds}},
  year          = {2017},
  month         = jan,
  number        = {1},
  pages         = {17},
  volume        = {834},
  adsurl        = {https://ui.adsabs.harvard.edu/abs/2017ApJ...834...17C},
  archiveprefix = {arXiv},
  doi           = {10.3847/1538-4357/834/1/17},
  eid           = {17},
  eprint        = {1603.08614},
  primaryclass  = {astro-ph.EP},
}

@Article{rowan-25b,
  author        = {{Rowan}, Connar and {Whitehead}, Henry and {Kocsis}, Bence},
  journal       = {\mnras},
  title         = {{Black hole merger rates in AGN: contribution from gas-captured binaries}},
  year          = {2025},
  month         = dec,
  number        = {4},
  pages         = {4576-4589},
  volume        = {544},
  adsurl        = {https://ui.adsabs.harvard.edu/abs/2025MNRAS.544.4576R},
  archiveprefix = {arXiv},
  doi           = {10.1093/mnras/staf1896},
  eprint        = {2412.12086},
  primaryclass  = {astro-ph.HE},
}

@Article{gwtc4-pop,
  author        = {{The LIGO Scientific Collaboration} and {the Virgo Collaboration} and {the KAGRA Collaboration}},
  journal       = {arXiv e-prints},
  title         = {{GWTC-4.0: Population Properties of Merging Compact Binaries}},
  year          = {2025},
  month         = aug,
  pages         = {arXiv:2508.18083},
  adsurl        = {https://ui.adsabs.harvard.edu/abs/2025arXiv250818083T},
  archiveprefix = {arXiv},
  doi           = {10.48550/arXiv.2508.18083},
  eid           = {arXiv:2508.18083},
  eprint        = {2508.18083},
  primaryclass  = {astro-ph.HE},
}

@ARTICLE{tyson-90,
       author = {{Tyson}, J.~A. and {Valdes}, F. and {Wenk}, R.~A.},
        title = "{Detection of Systematic Gravitational Lens Galaxy Image Alignments: Mapping Dark Matter in Galaxy Clusters}",
      journal = {\apjl},
         year = 1990,
        month = jan,
       volume = {349},
        pages = {L1},
          doi = {10.1086/185636},
       adsurl = {https://ui.adsabs.harvard.edu/abs/1990ApJ...349L...1T}
}

@ARTICLE{kaiser-93,
       author = {{Kaiser}, Nick and {Squires}, Gordon},
        title = "{Mapping the Dark Matter with Weak Gravitational Lensing}",
      journal = {\apj},
         year = 1993,
        month = feb,
       volume = {404},
        pages = {441},
          doi = {10.1086/172297},
       adsurl = {https://ui.adsabs.harvard.edu/abs/1993ApJ...404..441K}
}

@Article{dominik-sahu-00,
  author   = {{Dominik}, Martin and {Sahu}, Kailash C.},
  journal  = {\apj},
  title    = {{Astrometric Microlensing of Stars}},
  year     = {2000},
  month    = may,
  number   = {1},
  pages    = {213-226},
  volume   = {534},
  adsurl   = {https://ui.adsabs.harvard.edu/abs/2000ApJ...534..213D},
  doi      = {10.1086/308716},
}

@Article{hog-95,
  author   = {{Hog}, E. and {Novikov}, I.~D. and {Polnarev}, A.~G.},
  journal  = {\aap},
  title    = {{MACHO photometry and astrometry.}},
  year     = {1995},
  month    = feb,
  pages    = {287-294},
  volume   = {294},
  adsurl   = {https://ui.adsabs.harvard.edu/abs/1995A&A...294..287H},
}

@ARTICLE{miyamoto-yoshii-95,
       author = {{Miyamoto}, M. and {Yoshii}, Y.},
        title = "{Astrometry for Determining the MACHO Mass and Trajectory}",
      journal = {\aj},
         year = 1995,
        month = sep,
       volume = {110},
        pages = {1427},
          doi = {10.1086/117616},
       adsurl = {https://ui.adsabs.harvard.edu/abs/1995AJ....110.1427M}
}

@ARTICLE{walker-95,
       author = {{Walker}, Mark A.},
        title = "{Microlensed Image Motions}",
      journal = {\apj},
         year = 1995,
        month = nov,
       volume = {453},
        pages = {37},
          doi = {10.1086/176367},
       adsurl = {https://ui.adsabs.harvard.edu/abs/1995ApJ...453...37W}
}

@Article{bond-01,
  author        = {{Bond}, I.~A. and {Abe}, F. and {Dodd}, R.~J. and {Hearnshaw}, J.~B. and {Honda}, M. and {Jugaku}, J. and {Kilmartin}, P.~M. and {Marles}, A. and {Masuda}, K. and {Matsubara}, Y. and {Muraki}, Y. and {Nakamura}, T. and {Nankivell}, G. and {Noda}, S. and {Noguchi}, C. and {Ohnishi}, K. and {Rattenbury}, N.~J. and {Reid}, M. and {Saito}, To. and {Sato}, H. and {Sekiguchi}, M. and {Skuljan}, J. and {Sullivan}, D.~J. and {Sumi}, T. and {Takeuti}, M. and {Watase}, Y. and {Wilkinson}, S. and {Yamada}, R. and {Yanagisawa}, T. and {Yock}, P.~C.~M.},
  journal       = {\mnras},
  title         = {{Real-time difference imaging analysis of MOA Galactic bulge observations during 2000}},
  year          = {2001},
  month         = nov,
  number        = {3},
  pages         = {868-880},
  volume        = {327},
  adsurl        = {https://ui.adsabs.harvard.edu/abs/2001MNRAS.327..868B},
  archiveprefix = {arXiv},
  doi           = {10.1046/j.1365-8711.2001.04776.x},
  eprint        = {astro-ph/0102181},
  primaryclass  = {astro-ph},
}

@Article{chang-refsdal-79,
  author   = {{Chang}, K. and {Refsdal}, S.},
  journal  = {\nat},
  title    = {{Flux variations of QSO 0957 + 561 A, B and image splitting by stars near the light path}},
  year     = {1979},
  month    = dec,
  number   = {5739},
  pages    = {561-564},
  volume   = {282},
  adsurl   = {https://ui.adsabs.harvard.edu/abs/1979Natur.282..561C},
  doi      = {10.1038/282561a0},
}

@ARTICLE{16-martynov,
       author = {{Martynov}, D.~V. and others},
        title = "{Sensitivity of the Advanced LIGO detectors at the beginning of gravitational wave astronomy}",
      journal = {\prd},
         year = 2016,
        month = jun,
       volume = {93},
       number = {11},
          eid = {112004},
        pages = {112004},
          doi = {10.1103/PhysRevD.93.112004},
archivePrefix = {arXiv},
       eprint = {1604.00439},
 primaryClass = {astro-ph.IM},
       adsurl = {https://ui.adsabs.harvard.edu/abs/2016PhRvD..93k2004M}
}

@Article{branchesi-23,
  author        = {{Branchesi}, Marica and others},
  journal       = {\jcap},
  title         = {{Science with the Einstein Telescope: a comparison of different designs}},
  year          = {2023},
  month         = jul,
  number        = {7},
  pages         = {068},
  volume        = {2023},
  adsurl        = {https://ui.adsabs.harvard.edu/abs/2023JCAP...07..068B},
  archiveprefix = {arXiv},
  doi           = {10.1088/1475-7516/2023/07/068},
  eid           = {068},
  eprint        = {2303.15923},
  primaryclass  = {gr-qc},
}

@Article{pijnenburg-24,
  author        = {{Pijnenburg}, Martin and {Cusin}, Giulia and {Pitrou}, Cyril and {Uzan}, Jean-Philippe},
  journal       = {\prd},
  title         = {{Wave optics lensing of gravitational waves: Theory and phenomenology of triple systems in the LISA band}},
  year          = {2024},
  month         = aug,
  number        = {4},
  pages         = {044054},
  volume        = {110},
  adsurl        = {https://ui.adsabs.harvard.edu/abs/2024PhRvD.110d4054P},
  archiveprefix = {arXiv},
  doi           = {10.1103/PhysRevD.110.044054},
  eid           = {044054},
  eprint        = {2404.07186},
  primaryclass  = {gr-qc},
}

@Book{kepler-1609-first-second-laws,
  author  = {{Kepler}, Johannes},
  title   = {{Astronomia nova ..., seu physica coelestis, tradita commentariis de motibus stellae martis}},
  year    = {1609},
  adsurl  = {https://ui.adsabs.harvard.edu/abs/1609asno.book.....K},
  doi     = {10.3931/e-rara-558},
}

@BOOK{newton-1687-principia,
       author = {{Newton}, Isaac},
        title = "{Philosophiae Naturalis Principia Mathematica.}",
         year = 1687,
          doi = {10.3931/e-rara-440},
       adsurl = {https://ui.adsabs.harvard.edu/abs/1687pnpm.book.....N}
}

@Article{katz-18,
  author        = {{Katz}, Andrey and {Kopp}, Joachim and {Sibiryakov}, Sergey and {Xue}, Wei},
  journal       = {Journal of Cosmology and Astroparticle Physics},
  title         = {{Femtolensing by dark matter revisited}},
  year          = {2018},
  month         = dec,
  number        = {12},
  pages         = {005},
  volume        = {2018},
  adsurl        = {https://ui.adsabs.harvard.edu/abs/2018JCAP...12..005K},
  archiveprefix = {arXiv},
  doi           = {10.1088/1475-7516/2018/12/005},
  eid           = {005},
  eprint        = {1807.11495},
  primaryclass  = {astro-ph.CO},
}

\newpage

\end{document}